\documentclass[12pt]{article}

\usepackage[tbtags]{amsmath}
\usepackage{amsmath}
\usepackage{amssymb}
\usepackage{amsthm}

\newcommand{\ode}{ordinary differential equation}
\newcommand{\vf}{vector field}
\newcommand{\eq}{equivalence transformations}
\newcommand{\deq}{determining equation}
\newcommand{\gen}[1]{\partial_{#1}}

\newcommand{\pr}[1]{\rm pr^{(#1)}}
\newcommand{\curl}[1]{ \{#1\} }

\newcommand{\euclid}{\mathfrak e}

\newcommand{\R}{\mathbb{R}}

\newcommand{\nil}{\mathfrak {h}}
\newcommand{\semi}{\subset \hskip -3.8mm +}

\DeclareMathOperator{\Sl}{sl}

\DeclareMathOperator{\Su}{su}
\DeclareMathOperator{\So}{so}
\DeclareMathOperator{\Sp}{sp}

\DeclareMathOperator{\Di}{D}

\newtheorem{pro}{Proposition}
\newtheorem{thm}{Theorem}
\numberwithin{pro}{section}
\numberwithin{thm}{section}
\numberwithin{equation}{section}
\begin{document}
\title{\bf Symmetry classification of KdV-type nonlinear evolution
equations}
\author{F.~G\"{u}ng\"{o}r  \\ \small
Department of Mathematics, Faculty of Science and Letters,
\\
\small Istanbul Technical University, 80626, Istanbul, Turkey
\thanks{e-mail: gungorf@itu.edu.tr}\and \\
V.I.~Lahno \\
\small Pedagogical Institute, 2 Ostrogradskogo Street, 36003
Poltava, Ukraine
\thanks{e-mail: laggo@poltava.bank.gov.ua} \and
\\R.Z.~Zhdanov \\ \small Institute of Mathematics, 3
Tereshchenkivska Street, 252004 Kyiv, Ukraine\thanks{e-mail:
renat@imath.kiev.ua}}
\date{\today}
\maketitle
\begin{abstract}
Group classification of a class of third-order nonlinear evolution
equations generalizing KdV and mKdV equations is performed. It is
shown that there are two equations admitting simple Lie algebras
of dimension three.  Next, we prove that there exist only four
equations invariant with respect to Lie algebras having nontrivial
Levi factors of dimension four and six. Our analysis shows that
there are no equations invariant under algebras which are
semi-direct sums of Levi factor and radical. Making use of these
results we prove that there are three, nine, thirty-eight,
fifty-two inequivalent KdV-type nonlinear evolution equations
admitting one-, two-, three-, and four-dimensional solvable Lie
algebras, respectively. Finally, we perform a complete group
classification of the most general linear third-order evolution
equation.
\end{abstract}

\section{Introduction}
The purpose of this article is classifying equations of the form
\begin{equation}\label{main}
u_t = u_{xxx}  + F(x,t,u,u_x,u_{xx})
\end{equation}
which admit non-trivial Lie (point) symmetries. The standard Korteweg-de
Vries (KdV)  equation
$$u_t=u_{xxx}+u u_x$$
belongs to the family of evolution equations \eqref{main}.
Classification of the KdV equation with variable coefficients
(vcKdV)
\begin{equation}\label{vcKdV}
u_t=f(x,t)u u_x+g(x,t)u_{xxx}\qquad f\cdot g\ne 0
\end{equation}
by their symmetries is done in Ref. \cite{Gazeau92}, where
it is shown that the vcKdV can admit at most four-dimensional
Lie point symmetry group and those having four-dimensional symmetry
group can be transformed into the ordinary KdV equation  by local
point transformations. In Ref. \cite{Gungor96}, Eq.
\eqref{vcKdV} is investigated from the point of view of its integrability.
It is shown, in particular, that equations of the form \eqref{vcKdV}
with three-dimensional Lie point symmetry group have a property of
"partially integrability".

Our motivation is the same as for classifying vcKdV equations. We
start with a rather general class of nonlinear equations
generalizing \eqref{vcKdV} for $g_x=0$. Note that any $t$
dependent coefficient of $u_{xxx}$ in \eqref{main} can be
normalized by a reparametrization of time. The main advantage of
this classification is that, if we know the equation under study
admits a nontrivial symmetry group, then it is usually possible to
apply the whole spectrum of the methods and algorithms of Lie
group analysis. This enables us to derive exact analytical
solutions of equations under study, reveal their integrability
properties, find linearizing transformations, etc. Note that the
connection between Lie point symmetries and integrability was
discussed in \cite{Gungor96,Abellanas93}.

Recently, a novel generic approach to group classification of
low-dimensional partial differential equations (PDEs) has been
developed in \cite{Zhdanov99}. The full account of ideas and
algorithms applied can be found in the review paper
\cite{Basarab-Horwath01} where the approach in question has been
applied to classify the most general second-order quasi-linear
heat-conductivity equations admitting non-trivial Lie point
symmetries. Here we adopt the same approach which basically
consists of three steps. We first construct the equivalence group,
namely the most general group of point transformations that
transform any equation of the form \eqref{main} to a (possibly
different) equation belonging to the same class. Also, we find the
most general element of the symmetry group together with a
determining equation for $F$. As a second step, we realize
low-dimensional Lie algebras by vector fields of the above form up
to equivalence transformations. To this end, we use various
results on the structure of abstract Lie algebras
\cite{Mubarakzyanov63-1,Mubarakzyanov63-2, Mubarakzyanov66-3,
Patera76}. A review of the classification results of
non-isomorphic finite-dimensional Lie algebras can be found in
\cite{Basarab-Horwath01}. In the last step, after transforming
symmetry generators to canonical forms, we proceed to classifying
equations that admit nontrivial symmetries. We do this by
inserting these generators into the symmetry condition and solving
for $F$.

Let us mention that similar ideas have been used by Winternitz
and coworkers for group classification of several nonlinear
partial differential equations \cite{Gazeau92, Gagnon93,
Gungor01-3} and of discrete dynamical systems \cite{Levi96,
Gomez-Ullate99, Lafortune01}. Note also that group classification
of the nonlinear wave and Schr\"{o}dinger equations in the
same spirit has been done in \cite{Zhdanov000,Zhdanov00}.

The paper is organized as follows. In Section 2 we present the
determining equations for the symmetries and the equivalence group.
Section 3 is devoted to the classification of the equations invariant
under low-dimensional symmetry groups. In Section 4 we perform a
classification of linear equations in the class \eqref{main}. A
discussion of results and some conclusions are presented in the final
Section.

\section{Determining Equations and Equivalence Transformations}
Lie algebra of the symmetry group of equation
\eqref{main} is realized by \vf s of the form
\begin{equation}\label{vf}
X = \tau (x,t,u)\partial _t  + \xi (x,t,u)\partial _x  + \phi
(x,t,u)\partial _u.
\end{equation}

In order to implement the symmetry algorithm we need to calculate
the third order prolongation of the field \vf{} \eqref{vf} \cite{Olver86,
Ovsiannikov82, Bluman89}

\begin{equation}\label{pr}
{\pr{3}} X = X + \phi ^t \partial _{u_t }  + \phi ^x \partial
_{u_x } + \phi ^{xx} \partial _{u_{xx} }  + \phi ^{xxx} \partial
_{u_{xxx} }
\end{equation}
where
\[
\begin{gathered}
  \phi ^t  = D_t \phi  - u_t D_t \tau  - u_x D_t \xi , \hfill \\
  \phi ^x  = D_x \phi  - u_t D_x \tau  - u_x D_x \xi , \hfill \\
  \phi ^{xx}  = D_x \phi ^x  - u_{xt} D_x \tau  - u_{xx} D_x \xi , \hfill \\
  \phi ^{xxx}  = D_x \phi ^{xx}  - u_{xxt} D_x \tau  - u_{xxx} D_x \xi .
\hfill \\
\end{gathered}
\]
Here $D_x$ and $D_t$ denote the total space and time derivatives.
In order to find the coefficients of \vf{} we require that the
prolonged vector field \eqref{pr} annihilate equation \eqref{main} on
its solution manifold

\begin{equation}\label{symcon}
{\pr{3}}X(\Delta)\Bigl|_{\Delta=0}=0,\quad \Delta=u_t-u_{xxx}-F.
\end{equation}
Equating coefficients of linearly independent terms of invariance
criterion \eqref{symcon} to zero yields an overdetermined system of
linear PDEs (called determining equations). Solving this system we
obtain the following assertion.
\begin{pro}
The symmetry group of the nonlinear equation \eqref{main} for
an arbitrary (fixed) function $F$ is generated by the vector field
\begin{equation}\label{gvf}
X = \tau (t)\partial _t  + (\frac{{\dot \tau }} {3}x + \rho
(t))\partial _x  + \phi (x,t,u)\partial _u,
\end{equation}
where the functions $\tau(t)$, $\rho(t)$ and $\phi(x,t,u)$
satisfy the determining equation
\begin{equation}\label{deq}
\begin{array}{ll}
&-3\,u_{x}\,\dot{\rho} -
  x\,u_{x}\,\ddot{\tau}  -
  9\,u_{x}\,u_{xx}\,\phi _{uu} -
  3\,{u_{x}}^3\,\phi _{uuu} \\
&
  +3\,\phi _{t} -
  9\,u_{xx}\,\phi _{xu} -
  9\,{u_{x}}^2\,\phi _{xuu} -
  9\,u_{x}\,\phi _{xxu} -
  3\,\phi _{xxx} +
  3\left(\phi _{u} - \dot{\tau}
      \right)\,F\\
& +\left( 2\,u_{xx}\,\dot{\tau} -
     3\,u_{xx}\,\phi _{u} -
     3\,{u_{x}}^2\,\phi _{uu} -
     6\,u_{x}\,\phi _{xu} -
     3\,\phi _{xx} \right) \,F_{u_{xx}}\\
&
   + \left( u_{x}\,\dot{\tau} -
     3\,u_{x}\,\phi _{u} - 3\,\phi _{x}
     \right) \,F_{u_{x}} -
  3\,\phi \,F_{u} - 3\,\tau \,F_{t} -
  (3\,\rho \,+ x\, \dot{\tau})\,F_{x}=0.
\end{array}
\end{equation}
Here the dot over a symbol stands for time derivative.
\end{pro}
If there are no restrictions on $F$, then \eqref{deq} should be
satisfied identically, which is possible only when the symmetry
group is a trivial group of identity transformations. Here we
shall be concerned with the identification of all specific forms
of $F$ for which non-trivial symmetry groups occur. The basic idea
is to utilize the fact that for an arbitrarily fixed function $F$
all admissible vector fields form a Lie algebra. This immediately
implies the idea of using the classical results on classification
of low-dimensional Lie algebras obtained mostly in late sixties
\cite{Mubarakzyanov63-1,Mubarakzyanov63-2, Mubarakzyanov66-3}.
Saying it another way, we need to construct a kind of
representation theory on low-dimensional Lie algebras generated by
Lie vector fields preserving the manifold \eqref{deq}.

Our classification is up to equivalence under a group of
locally invertible point transformations

\begin{equation}\label{equ}
\tilde{t}=T(x,t,u),\quad \tilde{x}=Y(x,t,u),\quad
\tilde{u}=U(x,t,u),
\end{equation}
that preserve the form of the equation \eqref{main}, but
(possibly) change function $F$ into a new one, namely we have

\begin{equation}
\tilde u_{\tilde t}  = \tilde u_{\tilde x\tilde x\tilde x}  +
\tilde F(\tilde x,\tilde t,\tilde u,\tilde u_{\tilde x} ,\tilde
u_{\tilde x\tilde x} )
\end{equation}
Inserting \eqref{equ} into \eqref{main} and requiring
that the form of the equation be preserved, we arrive at the
following assertion.
\begin{pro}
The maximal equivalence group $\cal{E}$ has the form
\begin{equation}\label{eq}
\tilde t = T(t),\quad \tilde x = \dot T^{1/3} x + Y(t),\quad
\tilde u = U(x,t,u),
\end{equation}
where $\dot{T}\ne 0$, $U_u\ne 0$.
\end{pro}

We note that the Lie infinitesimal technique can also be used
to obtain the equivalence group \eqref{eq}. It is straightforward
to prove that the both approaches produce the same results.

We make use of \eq{} \eqref{eq} to transform \vf{} $X$ into
a convenient (canonical) form.

\begin{pro}\label{pro3}
Vector field \eqref{gvf} is equivalent within a point
transformation of the form \eqref{equ} to one of the following
vector fields:
\begin{equation}
X=\gen t,\quad X=\gen x,\quad X=\gen u.
\end{equation}
\end{pro}
\noindent{\bf Proof.} Transformation \eqref{eq} transform
\vf{} \eqref{gvf} into
\begin{equation}\label{trvf}
\begin{split}
X\to \tilde X &= \tau (t)\dot T(t)\partial _{\tilde t}  +
[\frac{1} {3}(\tau \dot T^{ - 1} \ddot T + \dot \tau )(\tilde x -
Y ) + \tau
\dot Y + \rho \dot T^{1/3} ]\partial _{\tilde x} \\
&+  [\tau U_t  + (\frac{1} {3}\dot \tau x + \rho )U_x  + \phi U_u
]\partial _{\tilde u}.
\end{split}
\end{equation}
There are two cases to consider:

I.) $\phi=0$. Choose $U=U(u)$ so that we have
\begin{equation}
\tilde X = \tau (t)\dot T(t)\partial _{\tilde t}  + [\frac{1}
{3}(\tau \dot T^{ - 1} \ddot T + \dot \tau )(\tilde x - Y) + \tau
\dot Y + \rho \dot T^{1/3} ]\partial _{\tilde x} \\
+   \phi U_u \partial _{\tilde u}.
\end{equation}
Now if $\tau=0,$ then $\rho\ne 0$ (otherwise $X$ would be zero) we
choose $T(t)$ to satisfy
$$\dot{T}=\rho^{-3}.$$  In this case $\tilde{X}$ is transformed into
$\gen{ \tilde{x}}$.

If $\tau\ne 0$, then we choose $T$ and $Y$ to satisfy
$$\dot{T}=\tau^{-1},\quad \tau \dot{Y}+\rho \dot{T}^{1/3}=0.$$
With this choice of $T$ and $Y$ vector field $\tilde{X}$ is transformed
into $\gen{\tilde{t}}$.

II.) $\phi\ne 0$. If $\tau=\rho=0$ then we can choose $U$ to
satisfy $\phi U_u=1$ so that we have $\tilde{X}=\gen {\tilde{u}}$.
Otherwise, $U$ can be chosen to satisfy
$$\tau U_t  + (\frac{1} {3}\dot \tau x + \rho )U_x  + \phi
U_u=0.$$ Hence we recover Case I.

Summing up, the vector field \eqref{gvf} is equivalent, up to
equivalence under $\cal{E}$, to one of the three standard vector
fields $\gen x, \gen t, \gen u$.
This completes the proof.

\section{Group classification of linear equations}
To the best of our knowledge no group classification of the most
general linear third-order PDE appears in literature. So we devote
this section to group classification of third-order PDEs
\begin{equation} \label{1}
u_t = f_1 (x,t) u_{xxx} +f_2(x,t) u_{xx} +f_3(x,t) u_x +f_4(x,t) u
+f_5(x,t).
\end{equation}

If we perform  the local change of variables $(x, t, u)\to
(\tilde{x}, \tilde{t}, \tilde{u})$ preserving the form of
\eqref{1}
\begin{equation}\label{trans}
\tilde t = t,\quad \tilde x = F(x,t), \quad u = V(x,t) v(\tilde x,
\tilde t) +G(x,t), \quad V \not =0, \ F_x \not =0
\end{equation}
we obtain
\begin{eqnarray*}
v_{\tilde t}  &=& f_1 F^3_x v_{\tilde x \tilde x  \tilde x} +\{3
f_1 V^{-1} [V_x
F^2_x +V F_x F_{xx}] + f_2 F^2_x \} v_{\tilde x \tilde x}+\\[.2cm]
&& \{f_1 V^{-1} [3 V_{xx} F_x +3 V_x F_{xx} +V F_{xxx} ]+f_2
V^{-1} [2 V_x
F_x+V f_{xx} ]+\\[.2cm]
&& f_3 F_x -F_t\} v_{\tilde x } +\{f_1 V^{-1} V_{xxx} +f_2 V^{-1} V_{xx}
+\\[.2cm]
&& +f_3 V^{-1} V_x +f_4 -V^{-1} V_t \} v+ V^{-1}[f_1 G_{xxx} +f_2
G_{xx} +f_3 G_x +f_4 G+f_5 -G_t].
\end{eqnarray*}
Now we choose the functions $F, V$, and $G$ in \eqref{trans} to
satisfy constraints
\begin{eqnarray*}
&& f_1 F^3_x =1, \\[.2cm]
&& G_t = f_1 G_{xxx} +f_2 G_{xx} +f_3 G_x +f_4 G +f_5, \\[.2cm]
&& 3 f_1 F^2_x V_x +[3 f_1 F_x F_{xx} +f_2 F^2_x ]V =0
\end{eqnarray*}
and thus normalize $f_1(x,t)\to 1$, and set $f_2(x,t)\to 0$,
$f_5(x,t)\to 0.$

Thus (\ref{1}) reduces to the following particular form:
\begin{equation}  \label{2}
u_t = u_{xxx} +A(x,t) u_x +B(x,t) u.
\end{equation}
Here $A, B$ are arbitrary smooth functions of $x$ and $t$.

The most general equivalence transformation preserving the class
of equations (\ref{2}), which is a subset of \eqref{eq}, reads as
\begin{equation}  \label{3}
\tilde t = T(t), \hskip 5mm \tilde x = \dot T^{{1}/{3}}x +Y(t),
\hskip 5mm \tilde u = V(t) u
\end{equation}
with $\dot T \not =0, \ V \not =0$.

Performing change of variables (\ref{3}) transforms equation
(\ref{2}) to become
\begin{equation}  \label{4}
\tilde u_{\tilde t} = \tilde u_{\tilde x \tilde x \tilde x}
+\tilde A \tilde u_{\tilde x} +\tilde B \tilde u,
\end{equation}
where the coefficients $\tilde A, \tilde B$ are expressed in terms
of the functions $A, B$ and their derivatives as follows
\begin{eqnarray} \label{5}
\tilde A &=& \dot T^{-1} (A \dot T^{\frac{1}{3}} -\frac{1}{3}
\ddot T \dot T^{-\frac{2}{3}} x - \dot Y), \\
\tilde B &=& \dot T^{-1} (B +V^{-1} \dot V). \nonumber
\end{eqnarray}

As equation (\ref{2}) is linear, it admits trivial
infinite-parameter group having the generator
$$
X{(\beta)} = \beta(x,t) \partial_u, \hskip 5mm \beta_t =
\beta_{xxx} +A \beta_x+B\beta_x
$$
and one-parameter group generated by the operator $u\partial_u$.
These symmetries give no non-trivial information about solution
structure of the equation under study and therefore are
neglected in the sequel.

Non-trivial invariance group of equation (\ref{2}) is generated by
operators of the form
\begin{equation} \label{6}
X = \tau(t) \partial_t +(\frac{1}{3} \dot \tau x +\rho (t))
\partial_x +\alpha(t)u\partial_u,
\end{equation}
functions $\tau, \rho, \alpha, A$ and $B$ satisfying equations
\begin{eqnarray} \label{7}
&&3 \dot \alpha -3 B \dot \tau -3 \tau B_t -B_x (3 \rho +x \dot
\tau) =0,
\\
&& -3 \dot \rho -x \ddot \tau -2 A \dot \tau -3 \tau A_t -A_x (3
\rho +x \dot \tau) =0. \nonumber
\end{eqnarray}

Provided $A = A(x,t), \ B = B(x,t)$ are arbitrary functions, $\tau
= \rho =0, \ \dot \alpha=0.$ So in this case equation (\ref{2}) admits
trivial symmetries only.

Transformation (\ref{3}) leaves operator $X_1 = u \partial_u$
invariant while transforming operator (\ref{4}) to become
\begin{equation} \label{8}
X {\mathop \rightarrow^{(\ref{3})} }\tilde X = \tau \dot T
\partial_{\tilde t} +[\tau (\frac{1}{3} \ddot T \dot
T^{-\frac{2}{3}} x +\dot Y) +\dot T^{\frac{1}{3}} (\frac{1}{3}
\dot \tau x +\rho)] \partial_{\tilde x}+(\tau \dot V +\alpha V) u
\partial_{\tilde u}.
\end{equation}

That is why, if $ \tau \not =0$ in (\ref{6}), then putting
$$\dot T = \tau^{-1},\quad \ Y = -\int^t \rho(\xi) \tau^{-\frac{4}{3}} (\xi)
d
\xi$$ and taking  $V$ as a non-zero solution of the equation
$$
\tau \dot V +\alpha V =0,
$$
in (\ref{3}) transforms (\ref{8}) to the canonical form of the
generator of time displacements
$$
\tilde X = \partial_{\tilde t}.
$$

Next, if $\tau =0, \ \rho \not =0$ in (\ref{6}), then putting
$\dot T = \rho^{-3}$ in (\ref{3}) yields the operator
$$
\tilde X = \partial_{\tilde x} +\alpha \tilde u \partial_{\tilde
u}.
$$

Finally, if $\tau = \rho =0,$ \ $\dot \alpha \not =0$ in
(\ref{6}), we put $T = \alpha$ in (\ref{3}) thus getting the
operator
$$
\tilde X = \tilde t \tilde u \partial_{\tilde u}.
$$

Taking into account the above considerations, we see that there
are transformations (\ref{3}), that transform operator (\ref{6})
to one of the following inequivalent forms:
$$
\partial_t,\quad \partial_x,\quad \partial_x
+f(t) u \partial_u \ (\dot f \not =0),\quad t u \partial_u.
$$

In what follows, we analyze each of the above operators
separately. \vspace{2mm}

\noindent \underline{Operator $X_1 = \partial_t$.} System of
determining equations (\ref{7}) for this operator reads as $$ B_t
= A_t =0, $$ whence it follows that $A = A(x), \ B = B(x).$
Inserting these functions into (\ref{7}) yields
\begin{eqnarray*}
&& 3 \dot \alpha - 3 B \dot \tau -B_x (3 \rho +x \dot \tau) =0, \\
&& -3 \dot \rho -x \ddot \tau -2 A \dot \tau -A_x (3 \rho+x \dot
\tau) =0.
\end{eqnarray*}

Analyzing the above system of ordinary differential equations
shows that for the case under consideration equation (\ref{2})
admits invariance group whose dimension is higher than one if and
only if
\begin{enumerate}
\renewcommand{\labelenumi}{\arabic{enumi})}

\renewcommand{\theenumi}{\arabic{enumi}}
\item $A = m x^{-2}, \ B = n x^{-3}, \ |m|+|n| \not =0$
with the additional symmetry operator $t \partial_t +\frac{1}{3} x
\partial_x;$
\item $A =0, \ B = \varepsilon x \ (\varepsilon = \pm 1)$ with the
additional symmetry operator $\partial_t +\varepsilon t u
\partial_u;$
\item $A = B \equiv 0$ with the additional symmetry operators $\partial_x, t
\partial_t +\frac{1}{3} x \partial_x.$
\end{enumerate}
\vspace{2mm}

\noindent \underline{Operator $X_2 = \partial_x$.} If equation
(\ref{2}) is invariant under $X_2$, then $A = A(t), \ B = B(t)$.
What is more, it follows from (\ref{5}) that there are
transformations (\ref{3}), which reduce equation (\ref{2}) to the
form (\ref{4}) with $\tilde A = \tilde B \equiv 0.$ So we arrive
at the already known case. \vspace{2mm}

\noindent \underline{Operator $X_3 =  \partial_x+f(t) u
\partial_u ,  \ (\dot f \not =0)$.} If equation (\ref{2})
admits operator $X_3$, then we have $A = 0, \ B = \dot f x.$
Inserting these expressions into (\ref{7}) yields
\begin{eqnarray} \label{9}
&& \dot \rho =0, \ \ \ddot \tau =0, \ \ \dot \alpha = \rho \dot f, \\
&& 3 \tau \ddot f +4 \dot \tau \dot f=0. \nonumber
\end{eqnarray}
From the first three equations it follows that $\rho = C_1, \ \tau
= C_2 t +C_3, \ \alpha = C_1 f +C_4, $ $C_1, C_2, C_3, C_4 \in
\mathbb{R}.$ Hence we conclude that the last equation of system
(\ref{9}) takes the form
$$
3(C_2 t +C_3) \ddot f +4 C_2 \dot f=0.
$$
Analyzing this equation we see that extension of the symmetry
algebra of equation (\ref{2}) with $A =0, \ B = \dot f x$ is only
possible when
\begin{eqnarray*}
f&=& 3 m t^{-\frac{1}{3}}, \ \ \ m \not =0;\\
f &= & \varepsilon t, \ \ \ \varepsilon = \pm 1.
\end{eqnarray*}
The second case has already been considered. In the first case the
basis of non-trivial invariance algebra is formed by the operators
$\partial_x +3 m t^{-\frac{1}{3}} u \partial_u, \ t \partial_t
+\frac{1}{3} x \partial_x.$ \vspace{2mm}

\noindent \underline{Operator $X_4 = t u \partial_u$.} Inserting
the coefficients of this operator into (\ref{7}) leads to the
contradiction $3 = 0$, whence it follows that the operator $X_4$
cannot be a symmetry operator of equation (\ref{2}).

We summarize the above classification results of in Table 1, where
we give the forms of the functions $A$ and $B$ and basis
operators of the non-trivial symmetry algebras of the
corresponding equations (\ref{2}).

\begin{table}\caption{Symmetry Classification of \eqref{2}}\label{tab1}
\begin{center}
\vskip 3mm
\begin{tabular}{|c|c|c|l|}
\hline
N   & $A$ & $B$ & Symmetry operators \\[2mm] \hline 1& $A(x)$ &
$B(x)$ &$\partial_t$ \\[2mm] \hline &&& \\ 2& $0$ & $\dot f (t)x$
&$\partial_x+f(t) u \partial_u, \ \dot f \not =0$
\\[2mm] \hline &&& \\ 3&  $m x^{-2}, \ m \in \mathbb{R}$ & $n x^{-3}, \ n
\in \mathbb{R},$ &$\partial_t, t \partial_t +\frac{1}{3} x
\partial_x$
\\[2mm] & & $ \ |m|+|n| \not =0$ & \\[2mm] \hline &&& \\ 4&  $0$ &
$\varepsilon x, \ \varepsilon = \pm 1$ &$\partial_t,  \partial_x
+\varepsilon t u
\partial_u$ \\[2mm] \hline &&& \\ 5&  $0$ & $-m
t^{-\frac{4}{3}}x,$ \   &$\partial_x+3m
t^{-\frac{1}{3}}u\partial_u,$ \\[2mm] & &$\ m \in \mathbb{R}, \ m \not
=0$&$t \partial_t +\frac{1}{3} x \partial_x$ \\[2mm] \hline &&& \\
6&  $a\in \mathbb{R}$ & $0$ &$\partial_t,  \partial_x, t\partial_t
+\frac{1}{3}
(x-2at)\partial_x$ \\[2mm] \hline
\end{tabular}
\end{center}
\end{table}

So the equation $u_t = u_{xxx}$ has the highest symmetry within
the class of equations (\ref{2}). Its maximal finite-dimensional
symmetry algebra is four-dimensional.

Note that according to \cite{Basarab-Horwath01} the class of
nonlinear equations of the form
\begin{equation}\label{10}
u_t = F(t,x,u,u_x) u_{xx} +G(t,x,u,u_x),\quad F \not =0,
\end{equation}
contains five nonlinear equations admitting five dimensional
symmetry algebras. Furthermore, an equation admitting
six-dimensional symmetry algebra is equivalent to the heat
equation. It is the linear heat conductivity equation
$u_t = u_{xx}$ that possess the largest symmetry group within the
class of second-order equations (\ref{10}).

This is not the case for the class of third-order PDEs under
consideration in the present paper. We shall see that there are
examples of nonlinear equations that admits higher symmetry
algebras than the linear equation does. For instance, the
nonlinear Schwarzian KdV equation \eqref{Schwarzian} admits a
six-dimensional symmetry algebra.

\section{Classification of equations invariant under semi-simple algebras
and
algebras having nontrivial Levi decompositions} In order to
describe equations (1.1) that admit Lie algebras isomorphic to the
Lie algebras having nontrivial Levi decomposition we need, first
of all, to describe equations whose invariance algebras are
semi-simple.

The lowest order semi-simple Lie algebras are isomorphic to one of
the following three-dimensional algebras:
\begin{eqnarray*}
\Sl(2, \mathbb{R})&:&  [X_1, X_3] = -2 X_2, \quad [X_1, X_2] =
X_1, \quad
[X_2, X_3] = X_3; \\
\So(3) &:&   [X_1, X_2] = X_3, \quad [X_2, X_3] = X_1, \quad [X_3,
X_1] = X_2.
\end{eqnarray*}

Taking into account our preliminary classification we conclude
that one of the basis operators  reduces to one of the canonical
forms $\partial_t, \partial_x, \partial_u$.

First, we study realizations of the algebra $\So(3)$ within the
class of operators (2.4).

Let $X_1 = \partial_t$ and let the operators $X_2, X_3$ be of the
form (2.4). Checking commutation relations $[X_1, X_2] = X_3, \
[X_3, X_1] = X_2$ we see that
\begin{eqnarray*}
X_2& =& 3 \alpha \cos t \partial_t +[-\alpha x \sin t +\beta \cos
(t+\gamma)] \partial_x+\varphi(x,u) \cos(t+\psi(x,u))
\partial_u,\\
X_3& =& -3 \alpha \sin t \partial_t -[\alpha x \cos t +\beta \sin
(t+\gamma)] \partial_x-\varphi(x,u) \sin(t+\psi(x,u)) \partial_u.
\end{eqnarray*}
Here $\alpha, \beta, \gamma$ are arbitrary real constants and
$\varphi$, $\psi$ are arbitrary real-valued smooth functions.

The third commutation relation $[X_2, X_3] = X_1$ implies that $9
\alpha^2 =-1.$ As this equation has no real solutions, there are
no realizations of $\So(3)$ with $X_1 = \partial_t$.

The same assertion holds for the cases when $X_1 = \partial_x$ and
$X_1 = \partial_u.$ So the class of operators (2.4) contains no
realizations of the algebra $\So(3)$. This means that there are no
$\So(3)$-invariant equations of the form (1.1).
\begin{thm}
There exist no realizations of the algebra $\So(3)$ in terms of
vector fields \eqref{gvf}. Hence no equation of the form
\eqref{main} is invariant under $\So(3)$ algebra.
\end{thm}

Similar reasoning yields that there are three inequivalent
realizations of the algebra $\Sl(2,\mathbb{R})$ by operators of
the form (2.4)
\begin{eqnarray*}
&& \{ \partial_t, t \partial_t +\frac{1}{3} x \partial_x, -t^2
\partial_t-\frac{2}{3} tx \partial_x \}, \\ && \{
\partial_t, t \partial_t +\frac{1}{3} x\partial_x, -t^2
\partial_t-\frac{2}{3} tx \partial_x-x^3 \partial_u \}, \\
&& \{ \partial_u, u \partial_u, -u^2 \partial_u \}.
\end{eqnarray*}

Inserting the coefficients of basis operators of the first
realization of the algebra $\Sl(2,\mathbb{R})$ into invariance
criterion yields the following classifying equations:
\begin{eqnarray*}
&& 2 u_{xx} F_{u_{xx}} +u_x F_{u_{x}} -x F_x -3 F =0, \\ && t(2
u_{xx} F_{u_{xx}} +u_x F_{u_x}-x F_x -3 F ) -x u_x =0,
\end{eqnarray*}
from which we get the equation $ x u_x =0$. Consequently, the
realization in question cannot be invariance algebra of the
equation under study.

The two remaining realizations of $\Sl(2,\mathbb{R})$ do yield
invariance algebras of equation (1.1). The forms of the function
$F$ in the corresponding invariant equations read as
\begin{eqnarray*}
&& \{ \partial_t, t \partial_t+\frac{1}{3} x \partial_x, -t^2
\partial_t-\frac{2}{3} tx \partial_x -x^3 \partial_u \}:\
F = -x^{-3} [2xu_x+\frac{1}{9} x^2 u^2_x -G(\omega_1,\omega_2)], \\
&& \ \ \ \ \ \ \ \ \omega_1 = 3u -x u_x, \ \ \omega_2 = 6u -x^2 u_{xx};\\
&& \{
\partial_u, u \partial_u, -u^2 \partial_u \} : F = -\frac{3}{2}
u^{-1}_x u^2_{xx} +u_x G(x,t).
\end{eqnarray*}

As any semi-simple or simple algebra contains either $\So(3)$ or
$\Sl(2,\mathbb{R})$ (or both) as sub\-algebra(s) \cite{Barut80},
the above result can be utilized to perform classification of
equations (1.1) admitting invariance algebras isomorphic to one
having a nontrivial Levi decomposition.

First we turn to the equation
\begin{equation} \label{2.1}
u_t = u_{xxx} -\frac{3}{2} u^{-1}_x u^2_{xx} +u_x G(x,t).
\end{equation}

Applying the Lie infinitesimal algorithm we see that
the maximal invariance algebra of equation (\ref{2.1})
is spanned by the operators $X_1 =
\partial_u, \ X_2 = u \partial_u, \ X_3 = -u^2 \partial_u,$ and
\begin{equation} \label{2.2}
X_4  = \tau (t) \partial_t +\Bigl (\frac{1}{3} \dot \tau x
+\rho(t)\Bigr )\partial_x,
\end{equation}
functions $\tau, \rho$ and $G$ satisfying the equation
\begin{equation} \label{2.3}
(x \dot{\tau} +3 \rho) G_x +3 \tau G_t +2 \dot{\tau} G +x
\ddot{\tau} +3 \dot{\rho} =0.
\end{equation}

By direct verification we ensure that the form of basis operators
of the realization of $\Sl(2,\mathbb{R})$ under study is not
altered by the transformations
\begin{equation} \label{2.4}
\tilde t = T(t), \ \ \tilde x = \dot T^{\frac{1}{3}} x +Y(t), \ \
\tilde u = \gamma u, \ \ \dot T \not =0, \ \ \gamma \not =0.
\end{equation}

As transformation (\ref{2.4}) reduces (\ref{2.2}) to the form
$$
X_4 {\mathop \rightarrow^{(\ref{2.4})} }\tilde {X_4} = \tau(t)
\dot T(t)
\partial_{\tilde t} +\frac{1}{3}(\tau \dot T^{-1} \ddot T
+\tau)(\tilde x-Y) +\tau \dot Y +\rho \dot T^{\frac{1}{3}}]
\partial_{\tilde x},
$$
we can put  $ X_4 =
\partial_t$ or $ X_4 = \partial_x$ within the equivalence
relation.

Provided $ X_4 = \partial_t,$ it follows from (\ref{2.3}) that $G
= \tilde G(x)$ in (\ref{2.1}). Next, if $ X_4 =
\partial_x$, then necessarily $G = \tilde G(t).$ Consequently,
class of equations (\ref{2.1}) contains two inequivalent equations
\begin{equation} \label{2.5}
u_t = u_{xxx} -\frac{3}{2} u^{-1}_x u^2_{xx} +u_x \tilde G(x)
\end{equation}
and
\begin{equation} \label{2.6}
u_t = u_{xxx} -\frac{3}{2} u^{-1}_x u^2_{xx} +u_x \tilde G(t)
\end{equation}
which are invariant under extensions of the algebra
$\Sl(2,\mathbb{R})$. Namely, they admit algebras
$\Sl(2,\mathbb{R}) \oplus \{ \partial_t \}$ and $\Sl(2,\mathbb{R})
\oplus \{
\partial_x \}$, correspondingly. What is more, if the
function $\tilde G = \tilde G(x)$ in (\ref{2.5}) is arbitrary, the
given algebra is maximal (in Lie sense) invariance algebra of equation
(\ref{2.5}).

Equation (\ref{2.6}) is reduced to PDE (\ref{2.5}) with $\tilde
G(x)=0$ with the help of the change of variables
$$ \tilde t = t, \ \ \ \tilde x =
x +\int^t \tilde{G}(\xi)\; d \xi, \ \ \ u = v(\tilde x, \tilde t).
$$
Therefore, we can restrict our further considerations to equation
(\ref{2.5}), where we need to differentiate between the cases
$\tilde G =0$ and $\tilde G \not =0$.

Classifying equation (\ref{2.3}) with $G = \tilde G(x)$ reads as
$$
(x \dot{\tau} +3 \rho) \tilde G_x +2 \dot{\tau} \tilde G +x
\ddot{\tau} +3 \dot{\rho} =0.
$$
Hence it follows that there are two cases providing for extension of
the symmetry algebra. Namely, the case when $\tilde G=0$, which
gives rise to two additional symmetry operators $X_5=t \partial_t
+\frac{1}{3} x \partial_x$ and $X_6=\partial_x$. Another case of
extension of symmetry of equation (\ref{2.5}) is when $\tilde G =
\lambda x^{-2} \ (\lambda \not =0)$. If this is the case,
(\ref{2.5}) admits the additional operator $X_5=t \partial_t
+\frac{1}{3} x \partial_x$.

Now we turn to the equation
\begin{eqnarray} \label{2.7}
u_t &=& u_{xxx} -2 x^{-2} u_x -\frac{1}{9} x^{-1} u^2_x +x^{-3}
G(\omega_1, \omega_2), \\ && \omega_1 = 3 u -x u_x, \ \ \omega_2 =
6u -x^2 u_{xx}. \nonumber
\end{eqnarray}

First of all, we ensure that the class of PDEs (\ref{2.7}) does
not contain equations, whose invariance algebras possess
semi-simple subalgebras of the dimension $n>3$.

It is a common knowledge \cite{Barut80}, that there are four types
of abstract simple Lie algebras over the field of real numbers:
\begin{itemize}
\item The type $A_{n-1} \ (n>1)$ contains four real forms of the
algebras $\Sl(n,\mathbb{C})$: $\Su(n)$, $\Sl(n,\mathbb{R})$,
$\Su(p,q)$ $(p+q=n, p \geq q), \Su^{*}(2n).$
\item  The type $B_n \ (n>1)$ contains two real forms of the
algebra $\So(2n+1, \mathbb{C}): $ $\So(2n+1)$, $\So(p,q)$ $(p+q
=2n+1, p>q).$
\item  The type $C_n \ (n\geq 1)$ contains three real forms of the
algebra $\Sp(n, \mathbb{C}): $ $\Sp(n)$, $\Sp(n,\mathbb{R})$,
$\Sp(p,q)$ $ (p+q =n, p\geq q).$
\item The type $D_n \ (n>1)$ contains three real forms of the
algebra $so(2n, \mathbb{C}):$ $\So(2n), \So(p,q)$ $(p+q=2n, \ p
\geq q), \So^{*} (2n).$
\end{itemize}

The lowest order classical semi-simple Lie algebras are
three-dimensional. The next admis\-sible dimension for classical
semi-simple Lie algebras is six. There are four non-isomorphic
semi-simple Lie algebras: $\So(4), \So(3,1), \So(2,2)$ and
$\So^{*}(4).$ As $\So(4) = \So(3) \oplus \So(3),\; \So^{*}(4) \sim
\So(3) \oplus \Sl(2,\mathbb{R}),$ and the algebra $\So(3,1)$
contains $\So(3)$ as a subalgebra, the algebra $\So(2,2)$ is the
only possible six-dimensional semi-simple algebra that might be
invariance algebra of equation (\ref{2.7}). Taking into account
that $\So(2,2) \sim \Sl(2,\mathbb{R}) \oplus \Sl(2,\mathbb{R})$
and choosing $\So(2,2) = \{ X_1, X_2, X_3 \} \oplus \{ {\tilde
X}_1, {\tilde X}_2, {\tilde X}_3 \}, $ where $X_1, X_2, X_3$ form
a basis of $\Sl(2,\mathbb{R})$, which is invariance algebra of
(\ref{2.7}) and ${\tilde X}_1, {\tilde X}_2, {\tilde X}_3 $ are of
the form (2.4), we require the commutation relations
$$
[X_i, {\tilde X}_j] =0 \ \ \ (i,j=1,2,3)
$$
to hold, whence
$$
{\tilde X}_j = \lambda_j \partial_u \ \ (j=1,2,3),
$$
where $\lambda_j$ are arbitrary real constants. Hence we conclude
that the class of operators (2.4) does not contain a realization
of $\So(2,2)$.

The same result holds for eight-dimensional semi-simple Lie
algebras $\Sl(3, \mathbb{R}), \Su(3),$ $\Su(2,1).$

As $\Su^{*}(4) \sim \So(5,1)$ and the algebra $\So(5,1)$ contains
$\So(4)$ as a subalgebra, the class of operators (2.4) contains no
realizations of $A_n$ and $D_n$ ($n>1$) type algebras that are
inequivalent to the algebra $\Sl(2,\mathbb{R})$.

The same assertion holds true for $B_n \ (n>1)$ and $C_n \ (n \geq
1)$ type Lie algebras. Indeed, $B_2$ type algebras contain
$\So(4)$ and $\So(3,1)$ and what is more,
$$
\Sp(2,\mathbb{R}) \sim \So(3,2)\supset \So(3,1),\quad \Sp(1,1)
\sim \So(4,1)\supset \So(4),\quad \Sp(2) \sim \So(5)\supset \So(4)
$$

What remains to be done is considering the exceptional semi-simple
Lie algebras that belongs to one of the following five types
\cite{Barut80}: $G_1, F_4, E_6, E_7, E_8.$ We consider in some
detail $G_1$ type Lie algebras.

The type $G_1$ contains one compact real form $g_2$ and one
non-compact real form $g'_2$. As $g_2 \cap g'_2 \sim \Su(2) \oplus
\Su(2) \sim \So(4)$ and the algebra $\So(4)$ has no realization
within the class of operators (2.4), the latter contains no
realizations of type $G_1$.

Summing up we conclude that class of PDEs (\ref{2.7})
contains no equations, whose invariance algebras are isomorphic to
$n$-dimensional semi-simple Lie algebras (or contains the latter
as subalgebras) under $n>3$.

Consider now equations (\ref{2.7}), whose invariance algebras has
non-trivial Levi factor. First, we turn to equations which are
invariant with respect to the Lie algebras that can be decomposed
into a direct sum of semi-simple Levi factor and radical,
$\Sl(2,\mathbb{R}) \oplus L,$ $L$ being a radical. To this end, we
will study possible extensions of the algebra $\Sl(2,\mathbb{R})$
by operators (2.4).

Let $\Sl(2,\mathbb{R}) = \{ X_1, X_2, X_3 \}, $ where $X_1, X_2,
X_3$, form a basis of the invariance algebra of equation
(\ref{2.7}). Then it follows from
$$
[X_i, Y]=0 \ (i=1,2,3),
$$
$Y$ being an operator of the form (2.4), that $Y = \lambda
\partial_u, \ \ \lambda =\text{const.}$ Hence $L$ is the one-dimensional
Lie algebra spanned by the operator $\partial_u.$ For equation
(\ref{2.7}) to admit the algebra $\Sl(2,\mathbb{R}) \oplus \{
\partial_u \}$ the equation
$$
G_{\omega _1 }  + 2G_{\omega _2 }  = 0,
$$
has to be satisfied, whence
$$
G = {\tilde G}({\sigma}), \ \ {\sigma} = x^2 u_{xx} -2 xu_x.
$$
Consequently, equation of the form (\ref{2.7}) admits invariance
algebra which is the direct sum of semi-simple Levi factor and
radical iff it reads as
\begin{equation} \label{p8}
u_t = u_{xxx} -2 x^{-2} u_x -\frac{1}{9} x^{-1} u^2_x+x^{-3}
{\tilde G}({\sigma}), \ \ {\sigma} = x^2 u_{xx} -2 x u_x.
\end{equation}

As equation (\ref{p8}) contains an arbitrary function of one
variable, we can perform direct group classification by
straightforward application of the Lie infinitesimal algorithm.
The determining equation for coefficients of the infinitesimal
symmetry operator are of the form
\begin{eqnarray*}
(a) && \phi_{uuu} =0; \\
(b) && 3 \phi_{uu} {\tilde G}_{\sigma}
+18 \phi_{uu} +9x \phi_{xuu}+\frac{1}{3} (x^{-1} \rho -\phi_u) =0;
\\
(c) && 6 x^{-1} (\phi_{xu} +x^{-2} \rho ){\tilde G}_{\sigma}
+9 x^{-2} {\sigma} \phi_{uu} +3 \rho_t +x \tau_{tt} + \\ && +6
x^{-1} (3 \phi_{xu} +2 x^{-2} \rho) +9 \phi_{xxu} -\frac{2}{3}
x^{-1} \phi_x =0; \\
(d)&& [-3 x^{-3} (\phi_u +2 x^{-1} \rho) {\sigma} +6 x^{-2} \phi_x
-3 x^{-1} \phi_{xx} ] {\tilde G}_{\sigma}+3 x^{-3} (\phi_u+3
x^{-1} \rho) {\tilde G}-
\\ && -9 x^{-2} \phi_{xu} {\sigma} +3[\phi_t -\phi_{xxx}
+2 x^{-2} \phi_x]=0.
\end{eqnarray*}
It follows from (a) that
\begin{equation} \label{p9}
\phi = f(x,t) u^2 +g(x,t) u +h(x,t),
\end{equation}
where $f, g, h$ are arbitrary smooth functions. Inserting
(\ref{p9}) into (b) yields
$$
6 f {\tilde G}_{\sigma} +36 f+18 xf_x +\frac{1}{3} (x^{-1}\rho -g)
-\frac{2}{3} f u =0.
$$
Taking into account that functions $f, \rho, g, {\tilde G}$ do not
depend on $u$, we get
$$
f=0, \ \ g = x^{-1} \rho.
$$
So that equation (c) reduces to
$$
3 \rho_t +x \tau_{tt} +12 x^{-3} \rho +\frac{2}{3} x^{-3} \rho u -
\frac{2}{3} x^{-1} h_x =0.
$$
Hence it follows that
$$
\rho =0, \ \ h = \frac{1}{2} x^{3} \tau_{tt} +\tilde h(t).
$$
Finally, inserting the obtained expression for $\varphi$ into
equation (d) gives
$$
\tau_{ttt}=0, \ \ \tilde h_t =0,
$$
whence
\begin{eqnarray*}
\tau&=& C_1 t^2 +C_2 t +C_3 , \\ \tilde h &=& C_4.
\end{eqnarray*}
Here $C_1, C_2, C_3, C_4$ are arbitrary (integration) constants.

Summing up, we conclude that the algebra $\Sl(2,\mathbb{R}) \oplus
\{
\partial_u \}$ is the maximal invariance algebra admitted by
equation (\ref{p8}). It cannot be extended by specifying the form
of an arbitrary function $\tilde G({\sigma}), \ \ {\sigma} = x^2
u_{xx} -2 x u_x$.

What remains to be done, is classifying equations (\ref{2.7}),
whose invariance algebras are iso\-mor\-phic to {\it semi-direct}
sums of semi-simple Levi factor and radical, i.e., whose
invariance algebras have the following structure:
$\Sl(2,\mathbb{R}) \subset \hskip -3.8mm + L.$ To perform this
classification we utilize the classification of these type of Lie
algebras obtained by Turkowski \cite{Turkowski88}.

We choose $\Sl(2,\mathbb{R}) = \{ {\rm v}_1, {\rm v}_2, {\rm v}_3
\}$ with
$$
{\rm v}_1=-2 t \partial_t-\frac{2}{3} x \partial_x, \ \ \  {\rm
v}_2=\partial_t, \ \ {\rm v}_3 =-t^2 \partial_t-\frac{2}{3} tx
\partial_x-x^3 \partial_u.
$$
According to \cite{Turkowski88}, there is only one
five-dimensional Lie algebra of the desired form
$\Sl(2,\mathbb{R}) \semi L$ with $L = \{ e_1, e_2 \}$, operators
$e_1, e_ 2$ satisfying the commutation relations:
\begin{eqnarray*}
&& [e_1, e_2] =0, \ \ [{\rm v}_1, e_1] = e_1, \ \ \ [{\rm v}_1,
e_2] = -e_2, \\ && [{\rm v}_2, e_1] =0, \ \ [{\rm v}_2, e_2] =
e_1, \\ &&  [{\rm v}_3, e_1] = e_2, \  \ \ \ [{\rm v}_3, e_2] = 0.
\end{eqnarray*}
as operators $e_1, e_2$ are necessarily of the form (2.4), we
easily get that
$$
e_1 = |x|^{-3/2} \partial_u, \quad e_2 = t |x|^{-3/2}
\partial_u.$$ However, checking the invariance criterion for the
above realization we find that the algebra in question cannot be
invariance algebra of an equation of the form (\ref{2.7}).

According to \cite{Turkowski88}, there exist three six-dimensional
Lie algebras that are semi-direct sums of semi-simple Levi factor
and radical, algebra $L$ being of the form $L = \{ e_1, e_2,
e_3\}$. Non-zero commutation relations for $e_1, e_2, e_3$ read as
\begin{eqnarray*}
1) && [{\rm v}_1, e_1] = 2 e_1, \ \ [{\rm v}_2, e_2 ] = 2 e_1, \ \
[{\rm v}_3, e_1] = e_2,\\ && [{\rm v}_1, e_3] = -2 e_3, \ \ [{\rm
v}_2, e_3 ] =  e_2, \ \ [{\rm v}_3, e_2] = 2e_3;\\ 2) && [{\rm
v}_1, e_1] =  e_1, \ \ [{\rm v}_2, e_2 ] = e_1, \ \ [{\rm v}_3,
e_1] = e_2,\\ && [{\rm v}_1, e_2] = - e_2, \ \ [e_1, e_2 ] =
e_3;\\ 3) && [{\rm v}_1, e_1] =  e_1, \ \ [{\rm v}_2, e_2 ] = e_1,
\ \ [{\rm v}_3, e_1] = e_2,\\ && [{\rm v}_1, e_2] = - e_2, \ \
[e_1, e_3 ] =  e_1, \ \ [e_2, e_3] = e_2.
\end{eqnarray*}

Solving the above relations we see that the corresponding
realizations cannot be invariance algebras of equation
(\ref{2.7}).

Next, we consider seven-dimensional algebras from the Turkowski's
classification. According to \cite{Turkowski88} there are five
inequivalent algebras of the targeted dimension. Four of them
contain the above five and six-dimensional algebras as
subalgebras. So we need to consider only the fifth algebra
$\Sl(2,\mathbb{R}) \subset \hskip -3.8mm + L,$ where $L = \{ e_1,
e_2, e_3, e_4 \}$ and the following commutation relations hold
\begin{eqnarray*}
&& [{\rm v}_1, e_1] = 3 e_1, \hskip 5mm [{\rm v}_2, e_2] = 3 e_1,
\\ && [{\rm v}_3, e_1] =  e_2, \hskip 7mm [{\rm v}_1, e_2] =  e_2,
\\ && [{\rm v}_2, e_3] = 2 e_2, \hskip 5mm [{\rm v}_3, e_2] = 2
e_3, \\ && [{\rm v}_1, e_3] = - e_3, \hskip 5mm [{\rm v}_2, e_4] =
e_3, \\ && [{\rm v}_3, e_3] = 3 e_4, \hskip 5mm [{\rm v}_1, e_4] =
-3 e_4.
\end{eqnarray*}
The most general form of operators $e_1, e_2, e_3, e_4$ satisfying
the above relations is as follows:
\begin{eqnarray*}
e_1 &=& |x|^{-\frac{9}{2}} \partial_u, \hskip 10mm e_2 = 3 t
|x|^{-\frac{9}{2}} \partial_u, \\ e_3 &=& 3 t^2 |x|^{-\frac{9}{2}}
\partial_u, \hskip 5mm e_4 =  t^3 |x|^{-\frac{9}{2}} \partial_u.
\end{eqnarray*}
However, verifying the invariance criterion yields that this
algebra cannot be symmetry algebra of equation (\ref{2.7}).

Thus we proved that the class of PDEs (\ref{2.7}) contains no
equations admitting symmetry algebras of the dimension $n\leq 7$,
which are semi-direct sums of Levi factor and radical. It is
natural to conjecture that the same assertion holds for an arbitrary
$n$. To prove this fact we need to consider in full details
classification of nonlinear equations (1.1), whose invariance
algebras are solvable.

Let us sum up the above results as theorems.
\begin{thm}
The class of PDEs (1.1) contains two inequivalent equations whose
invariance algebra are semi-simple ($\Sl(2,\mathbb{R})$)
\begin{eqnarray*}
u_t &=& u_{xxx} -\frac{3}{2} u^{-1}_x u^2_{xx} +u_x G(x,t); \\ u_t
&=& -x^{-3} [2 x u_x +\frac{1}{9} x^2 u^2_x -G(\omega_1,\omega_2)], \\
&& \omega_1 = 3 u -x u_x, \ \ \omega_2 = 6 u -x^2 u_{xx}.
\end{eqnarray*}
The maximal invariance algebras of the above equations under
arbitrary $G$ read as
\begin{eqnarray*}
\Sl^1(2,\mathbb{R}) &=& \{ \partial_u, u \partial_u, -u^2
\partial_u \}; \\ \Sl^2(2,\mathbb{R}) &=& \{ \partial_t,
t
\partial_t+\frac{1}{3} x \partial_x, -t^2 \partial_t-\frac{2}{3} t
x \partial_x- x^3\partial_u \}.
\end{eqnarray*}
\end{thm}
\begin{thm}
Nonlinear equation (1.1) whose invariance algebra is isomorphic to
a Lie algebra having non-trivial Levi decomposition is represented
by one of the following equations:
\begin{align}
&u_t = u_{xxx} -\frac{3}{2} u^{-1}_x u^2_{xx} +u_x {\tilde G}(x),
\quad  \Sl^1(2,R) \oplus \{
\partial_t \};\\
&u_t = u_{xxx} -\frac{3}{2}u^{-1}_x u^2_{xx} +\lambda x^{-2} u_x,
\ \ \lambda \not =0, \quad \Sl^1 (2, \mathbb{R}) \oplus \{
\partial_t, t
\partial_t +\frac{1}{3} x \partial_x \};\\
&u_t = u_{xxx} -\frac{3}{2} u^{-1}_x u^2_{xx}, \quad \Sl^1
(2,\mathbb{R}) \oplus \{ \partial_t, \partial_x, t
\partial_t+\frac{1}{3} x
\partial_x \};\label{semi}\\
&u_t = u_{xxx} -2 x^{-2} u_x -\frac{1}{9} x^{-1} u^2_x +x^{-3}
{\tilde G}({\sigma}), \ {\sigma} = x^2 u_{xx} -2 x u_x, \quad
\Sl^2 (2,\mathbb{R}) \oplus \{\partial_u \},
\end{align}
where $\tilde G$ is an arbitrary function of $x$ or $\sigma$.
Moreover, the associated  symmetry algebras are maximal.
\end{thm}
Note that  equation \eqref{semi} can be expressed in the form
\begin{equation}\label{Schwarzian}
\frac{u_t}{u_x}=\curl{u;x},
\end{equation}
where $\curl{u;x}$ denotes the Schwarzian derivative of $u$ with
respect to $x$. It is known that a non-point transformation taking
this equation into the usual KdV  exists.

\section{Classification of equations invariant under low-dimensional
solvable symmetry algebras}

In this section we apply the strategy summarized in the Introduction
to identify representative classes of equations of the form \eqref{main}
invariant under one-, two-, and three-dimensional solvable symmetry
algebras. In order to approach this task in a systematic manner
we realize all possible inequivalent algebras in terms of vector
fields \eqref{gvf} under the action of the equivalence group $\cal{E}$.

\subsection{Equations with one-dimensional symmetry algebras}
We assume that for a given $F$, equation \eqref{main} is invariant
under  a one-parameter symmetry group, generated by the \vf{}
\eqref{gvf} with coefficients subject to the constraint
\eqref{deq}. We make use of Proposition  \ref{pro3}
which characterizes the canonical forms of the vector field $X$ of
\eqref{gvf}. We then substitute the coefficients of the canonical
\vf{} into the \deq{} \eqref{deq}, which is a first order linear
homogeneous PDE for $F$, and solve the latter in order to
construct invariant equations.

According to Proposition  \ref{pro3} we have three types of
one-dimensional symmetry algebras

\begin{equation}\label{one-dim}
A_{1,1}: X_1=\gen t,\quad A_{1,2}: X_1=\gen x,\quad A_{1,3}:
X_1=\gen u.
\end{equation}
The corresponding invariant equations will have the form
\begin{subequations}\label{one-dim-F}
\begin{eqnarray}
A_{1,1} &:\quad u_t  = u_{xxx}  + F(x,u,u_x ,u_{xx} ),\\
A_{1,2} &:\quad u_t  = u_{xxx}  + F(t,u,u_x ,u_{xx} ),\\
A_{1,3} &:\quad u_t  = u_{xxx}  + F(x,t,u_x ,u_{xx} ).
\end{eqnarray}
\end{subequations}

\begin{thm}
There are  three inequivalent classes of equations \eqref{main}
invariant under one-parameter symmetry group.  Their
representatives are given by \eqref{one-dim-F}.
\end{thm}

\subsection{Equations with two-dimensional symmetry algebras}
There are two isomorphy classes of two-dimensional Lie algebras,
abelian  and non-abelian satisfying the commutation relations
$[X_1,X_2]=\kappa X_2$, $\kappa=0,1.$ We denote them by $A_{2,1}$
and $A_{2,2}$.

{\bf I. Abelian:}

\noindent We start from each of the  one-dimensional cases
obtained in \eqref{one-dim} and add to it \vf s $X_2$ of the form
\eqref{gvf} commuting with $X_1$. We then simplify $X_2$ by \eq{}
leaving the \vf{} $X_1$ invariant. For further details we refer
the reader to Ref. \cite{Zhdanov99}.  The standardized $X_2$ and
the restricted form of $F$ in \eqref{one-dim-F} are then
substituted into \eqref{deq}. Solving this equation will further
restrict the form of the function $F$. The number of variables of
$F$ reduces by one, three variables in this case. Thus, we find
that there exist precisely four classes of two-dimensional abelian
symmetry algebras represented by the following ones:

\begin{eqnarray}\label{two-dim-1}
A_{2,1}^{1}&: &X_1=\gen t,\quad X_2=\gen x,\quad F=F(u,u_x,u_{xx})\\
A_{2,1}^{2}&: &X_1=\gen t,\quad X_2=\gen u,\quad F=F(x,u_x,u_{xx})\\
A_{2,1}^{3}&: &X_1=\gen x,\quad X_2=\alpha(t)\gen x+\gen u,\quad
F=-\dot{\alpha}u u_x+\tilde F(t,u_x,u_{xx})\\
A_{2,1}^{4}&: &X_1=\gen u,\quad X_2=g(x,t)\gen u ,\quad g_x\ne
\mbox{const.}, \\
&& F=(g_t-g_{xxx})g_x^{-1}u_x+\tilde F(x,t,\omega),\quad
\omega=g_{xx}u_x-g_x u_{xx}.\nonumber
\end{eqnarray}

{\bf II. Non-abelian:}

Imposing that $X_1$ reads as \eqref{one-dim} and $X_2$ is in generic form
and that they satisfy $[X_1,X_2]=X_2$, we find that five classes of symmetry
algebras exist. Those algebras and nonlinear functions $F$ are represented
by

\begin{eqnarray*}\label{two-dim-2}
A_{2,2}^{1}&: &X_1=\gen t,\quad X_2=-t\gen t-\frac{x}{3}\gen x,\\
&& F=x^{-3}\tilde F(u,\omega_1,\omega_2),\quad \omega_1=x
u_x,\quad \omega_2=x^2 u_{xx},\nonumber\\
A_{2,2}^{2}&: &X_1=-3t\gen t-x\gen x,\quad X_2=\gen x,\\
&& F=t^{-1}\tilde F(u,\omega_1,\omega_2),\quad
\omega_1=t^{1/3}u_x,\quad
\omega_2=t^{2/3}u_{xx},\nonumber\\
A_{2,2}^{3}&: &X_1=-u\gen u,\quad X_2=\gen u,\quad F=u_x
\tilde F(x,t,\omega),\quad \omega=u_x^{-1}u_{xx},\\
A_{2,2}^{4}&: &X_1=\gen x-u\gen u,\quad X_2=\gen u,\\
&& F=e^{-x}\tilde F(t,\omega_1,\omega_2),\quad
\omega_1=e^{x}u_{x},\quad \omega_2=e^{x}u_{xx},\nonumber\\
A_{2,2}^{5}&: &X_1=\gen t-u\gen u,\quad  X_2=\gen u,\\
&& F=u_x\tilde F(x,\omega_1,\omega_2),\quad
\omega_1=e^{t}u_{x},\quad \omega_2=e^{t}u_{xx}.\nonumber
\end{eqnarray*}

\begin{thm}
There exist nine classes of two-dimensional symmetry algebras
admitted by equation \eqref{main}. They are represented by the algebras
$A_{2,1}^1,\ldots,A_{2,1}^4$ and $A_{2,2}^1,\ldots,A_{2,2}^5$.
\end{thm}

\subsection{Equations with three-dimensional symmetry algebras}

{\bf Decomposable algebras:}

A Lie algebra is decomposable if it can be written as a direct sum
of two or more Lie algebras $L=L_1\oplus L_2$ with $[L_1,L_2]=0$.
There are two types of 3-dimensional  decomposable Lie algebras,
$A_{3,1}=3A_1=A_1\oplus A_2\oplus A_3$ with $[X_i,X_j]=0$ for
$i,j=1,2,3$ and $A_{3,2}=A_{2,2}\oplus A_1$ with $[X_1,X_2]=X_2$,
$[X_1,X_3]=0$, $[X_2,X_3]=0$.

We start from the two-dimensional algebras in \eqref{two-dim-1}
and add a further linearly independent \vf{} $X_3$ in the form
\eqref{gvf} and impose the above commutation relations. We
simplify $X_3$ using \eq{} leaving the space $\curl{X_1, X_2}$
invariant. We present the following result without proof. We
emphasize that there exist several realizations that do not
produce invariant equations of the form \eqref{main}.
\begin{eqnarray*}\label{three-dim}
A_{3,1}^1&: &X_1=\gen t,\quad X_2=\gen x,\quad X_3=\gen u,\quad
F=F(u_x,u_{xx}),\\
A_{3,1}^2&: &X_1=\gen t,\quad X_2=\gen u,\quad X_3=f(x)\gen u,\\
&& F=-\frac{f'''}{f'}u_x+\tilde F(x,\omega),\quad \omega=f''
u_x-f'
u_{xx},\nonumber\\
A_{3,2}^1&: &X_1=-t\gen t-\frac{x}{3}\gen x,\quad X_2=\gen t,\quad
X_3=\gen u,\\
&& F=x^{-3}\tilde F(\omega_1,\omega_2),\quad \omega_1=x u_x,\quad
\omega_2=x^2 u_{xx},\nonumber\\
A_{3,2}^2&: &X_1=-3t\gen t-x\gen x,\quad X_2=\gen x,\quad X_3=\gen u,\\
&& F=t^{-1}\tilde F(\omega_1,\omega_2),\quad
\omega_1=tu_{x}^3,\quad
\omega_2=t^2u_{xx}^3, \\
A_{3,2}^3&: &X_1=-3t\gen t-x\gen x,\quad X_2=\gen x,\quad
X_3=t^{1/3}\gen x+\gen u,\\
&& F=-\frac{1}{3}t^{-2/3}uu_x+t^{-1}\tilde
F(\omega_1,\omega_2),\quad
\omega_1=tu_{x}^3,\quad \omega_2=t^2u_{xx}^3, \\
A_{3,2}^4&: &X_1=\gen x-u\gen u,\quad  X_2=\gen u,\quad
X_3=e^{-x}f(t)\gen u,\quad f\ne 0,\\
&& F=-(1+\frac{\dot{f}}{f})u_x+e^{-x}\tilde F(t,\omega),\quad
\omega=e^x(u_{x}+u_{xx}),\\
A_{3,2}^5&: &X_1=\gen x-u\gen u,\quad  X_2=\gen u,\quad
X_3=\alpha(t)\gen x,\quad {\alpha}\ne 0,\\
&& F=-\frac{\dot\alpha}{\alpha}u_x\ln(e^x
u_{x})+u_x\tilde F(t,\omega),\quad \omega=u_{x}^{-1}u_{xx},\\
A_{3,2}^6&: &X_1=\gen x-u\gen u,\quad  X_2=\gen u,\quad
X_3=\gen t,\\
&& F=e^{-x}\tilde F(\omega_1,\omega_2),\quad
\omega_1=e^x u_{x},\quad \omega_2=e^x u_{xx},\\
\end{eqnarray*}
\begin{eqnarray*}
A_{3,2}^7&: &X_1=\gen t-u\gen u,\quad  X_2=\gen u,\quad
X_3=e^{-t} f(x)\gen u,\quad f'\ne 0\\
&& F=-\frac{f'''+f}{f'}u_x+e^{-t}\tilde F(x,\omega),\quad
\omega=e^t(f'' u_x-f' u_{xx}),\\
A_{3,2}^8&: &X_1=\gen t-u\gen u,\quad  X_2=\gen u,\quad
X_3=\gen x,\\
&& F=e^{-t}\tilde F(\omega_1,\omega_2),\quad \omega_1=e^t
u_{x},\quad \omega_2=e^t u_{xx},\\
A_{3,2}^9&: &X_1=\gen t-u\gen u,\quad  X_2=\gen u,\quad
X_3=\gen t+\lambda \gen x,\quad \lambda \ne 0,\\
&& F=\exp{(x/t-\lambda)}\tilde F(\omega_1,\omega_2),\\
&&\omega_1=\exp{(t-x/\lambda)}u_x,\quad
\omega_2=\exp{(t-x/\lambda)}u_{xx},\\
A_{3,2}^{10}&: &X_1=\gen t-u\gen u,\quad  X_2=\gen u,\quad
X_3=\gen t,\\
&& F=u_{x}\tilde F(x,\omega),\quad \omega={u_{x}}^{-1}{u_{xx}}.
\end{eqnarray*}

{\bf Nondecomposable algebras:} The isomorphy classes of these
algebras are represented by the following list:



\begin{eqnarray*}
A_{3,3} &:& [X_2,X_3]=X_1,\quad [X_1,X_2]=[X_1,X_3]=0;  \\
A_{3,4} &:& [X_1, X_3] = X_1, \quad [X_2, X_3] = X_1+X_2; \\
A_{3,5} &:& [X_1, X_3] = X_1, \quad [X_2, X_3] = X_2; \\
A_{3,6} &:& [X_1, X_3] = X_1, \quad [X_2, X_3] = -X_2; \\
A_{3,7} &:& [X_1, X_3] = X_1, \quad [X_2, X_3] = q X_2 \;
(0<|q|<1) ;\\
A_{3,8} &:& [X_1, X_3] = -X_2, \quad [X_2, X_3] = X_1;\\
A_{3,9} &:& [X_1, X_3] = q X_1-X_2, \quad [X_2, X_3] = X_1+qX_2,
\; q>0.
\end{eqnarray*}
{\bf Remark:} Solvable nondecomposable algebras can be written
as semidirect sums of a one-dimensional subalgebra $\curl{X_3}$
and an abelian ideal $\curl{X_1, X_2}$. Note that the algebras
$A_{3,6}$ and $A_{3,8}$ are isomorphic to $\euclid(1,1)$, and
$\euclid(2)$, respectively. The algebra $A_{3,3}$ is a nonabelian
nilpotent algebra (Heisenberg algebra).

The commutation relations of the algebras in question
can be represented in the matrix notation
\[
\left( {\begin{array}{*{20}c}
   {[X_1 ,X_3 ]}  \\
   {[X_2 ,X_3 ]}  \\

\end{array} } \right) = J\left( {\begin{array}{*{20}c}
   {X_1 }  \\
   {X_2 }  \\

\end{array} } \right),\quad   [X_1,X_2]=0
\]
where $J$ is a $2\times 2$ real matrix that can be taken in Jordan
canonical form.

A solvable three-dimensional Lie
algebra always possesses a two-dimensional abelian ideal. We assume
that the ideal $\curl{X_1,X_2}$ is already of the form
\eqref{two-dim-1} and add a third element $X_3$ in the form
\eqref{gvf} acting on the ideal. Imposing commutation relations
and simplifying with equivalence transformations \eqref{equ} (we
consider each canonical form of the matrix individually) yield the
realizations of solvable Lie algebras together with the
corresponding invariant equations.

There exist nine classes of realizations of nilpotent
algebras which give rise to invariant equations.

\begin{equation}\label{3dim-solv-3}
A_{3,3}  : \quad J = \left( {\begin{array}{*{20}c}
   0 & 0  \\
   1 & 0  \\
\end{array} } \right).
\end{equation}

\begin{eqnarray*}
A_{3,3}^{1}&: &X_1=\gen t,\quad X_2=\gen u,\quad
X_3=t\gen u+\lambda \gen x,\quad \lambda>0\\
&& F=\frac{x}{\lambda}+\tilde F(u_x,u_{xx}),\\
A_{3,3}^{2}&: &X_1=\gen u,\quad X_2=\gen x,\quad
X_3=x\gen u+b(t) \gen x,\quad \dot{b}\ne0\\
&& F=-\frac{\dot{b}}{2}u_{x}^2+\tilde F(t,u_{xx}),\\
A_{3,3}^{3}&: &X_1=\gen u,\quad X_2=\gen x,\quad
X_3=x\gen u+\lambda \gen t,\quad \lambda\ne0\\
&& F=\tilde F(t-3\lambda u_x,u_{xx}),\\
A_{3,3}^{4}&: &X_1=\gen u+3\lambda t^{1/2}\gen x,\quad X_2=\gen
x,\\
&& X_3=6\lambda t^{3/2}\gen t+3\lambda t^{1/2}x \gen
x+(x-3\lambda t^{1/2} u)\gen u
,\quad \lambda\ne 0\\
&& F=-\frac{3}{2}\lambda t^{-1/2}u u_x+t^{-2}\tilde
F(\omega_1,\omega_2),\quad
\omega_1=tu_x-\frac{1}{3\lambda}t^{1/2},\quad
\omega_2=t^{3/2}u_{xx},\\
A_{3,3}^{5}&: &X_1=\gen x,\quad X_2=\gen t,\quad X_3=t\gen x+\gen u,\\
&& F=-uu_x+\tilde F(u_x, u_{xx}),\\
A_{3,3}^{6}&: &X_1=\gen u,\quad X_2=(f(x)-t)\gen u,\quad X_3=\gen t,\;(f'\ne
0)\\
&& F=-(1+f''')(f')^{-1} u_x +\tilde F(x, \omega), \quad \omega =
f''u_x-f'u_{xx},\\
A_{3,3}^{7}&: &X_1=\gen u,\quad X_2=(t-x)\gen u,\quad X_3=\gen x,\\
&& F=u_x+\tilde F(t, u_{xx}),\\
A_{3,3}^{8}&: &X_1=\gen u,\quad X_2=-x\gen u,\quad X_3=\gen x,\\
&& F=-\tilde F(t, u_{xx}),\\
A_{3,3}^{9}&: &X_1=-x^{-1}\gen u,\quad X_2=\gen u,\quad
X_3=\gen x-x^{-1}u\gen u,\\
&& F=3x^{-1}u_{xx}+x^{-1}\tilde F(t, \omega),\quad
\omega=2u_x+xu_{xx}.
\end{eqnarray*}

\begin{equation}\label{3-dim-solv-4}
A_{3,4}  : \quad J = \left( {\begin{array}{*{20}c}
   1 & 0  \\
   1 & 1  \\
\end{array} } \right).
\end{equation}

\begin{eqnarray*}
A_{3,4}^{1}&: &X_1=\gen u,\quad X_2=\gen t,\quad
X_3=t\gen t+\frac{x}{3} \gen x+(u+t)\gen u,\\
&& F=3\ln x+\tilde F(\omega_1,\omega_2),\quad
\omega_1=x^{-2}u_x,\quad
\omega_2=x^{-1}u_{xx},\\
A_{3,4}^{2}&: &X_1=\gen x,\quad X_2=\gen u-\frac{1}{3}\ln t\gen x,\quad
X_3=3t\gen t+x\gen x+u\gen u,\\
&& F=\frac{1}{3t}u u_{x}+t^{-2/3}\tilde F(u_x,\omega),\quad
\omega=t^{1/3}u_{xx},\\
A_{3,4}^{3}&: &X_1=\gen u,\quad X_2=\gen x,\quad
X_3=3t\gen t+x\gen x+(u+x)\gen u,\\
&& F=t^{-2/3}\tilde F(\omega_1,\omega_2),\quad
\omega_1=u_x-\frac{1}{3}\ln t,\;\omega_2=t^{1/3}u_{xx},\\
A_{3,4}^4 &:& X_1 = \alpha(t)\partial_x+\partial_u, \ X_2 = \partial_x,\\
&&  X_3 = (\alpha')^{-1} \alpha^2  \partial_t+(1+\alpha)x\partial_x+
[x+(1-\alpha) u]\partial_u, \ \ \alpha' \not =0,\\
&& \alpha^2 \alpha'' +(3+\alpha)(\alpha')^2 =0, \\
&& F = -\alpha' u u_x +\alpha^{-4}\exp(2\alpha^{-1})
\tilde F(\omega_1, \omega_2), \\
&& \omega_1 = \alpha^3\exp(-\alpha^{-1}) u_{xx},
\quad
\omega_2 = \alpha^2 u_x -\alpha,\\
A_{3,4}^{5}&: &X_1=\gen u,\quad X_2=(-t+f(x))\gen u,\quad
X_3=\gen t+u\gen u,\quad f'\ne 0,\\
&& F=-(1+f''')(f')^{-1}u_x+e^t\tilde F(x,\omega),\quad
\omega=e^{-t}(f''u_x-f'u_{xx}),\\
A_{3,4}^{6}&: &X_1=\gen u,\quad X_2=-x\gen u,\quad
X_3=\gen x+u\gen u,\\
&& F=e^x\tilde F(t,\omega),\quad \omega=e^{-x}u_{xx}.\\
\end{eqnarray*}

\begin{equation}\label{3-dim-solv-5}
A_{3,5} : \quad J = \left( {\begin{array}{*{20}c}
   1 & 0  \\
   0 & 1  \\
\end{array} } \right).
\end{equation}

\begin{eqnarray*}
A_{3,5}^{1}&: &X_1=\gen t,\quad X_2=\gen u,\quad
X_3=t\gen t+\frac{x}{3} \gen x+u\gen u,\\
&& F=\tilde F(\omega_1,\omega_2),\quad \omega_1=x^{-2}u_{x},\;
\omega_2=x^{-1}u_{xx},\\
A_{3,5}^{2}&: &X_1=\gen x,\quad X_2=\gen u,\quad
X_3=3t\gen t+{x}\gen x+u\gen u,\\
&& F=t^{-2/3}\tilde F(u_{x},t^{1/3}u_{xx}),\\
A^3_{3.5} &:& X_1 = \partial_u, \ X_2= f(x) \partial_u, X_3 =
\partial_t +u \partial_u, \ f' \not =0,\\ && F = -f'''(f')^{-1} u_x
+e^t \tilde F(x,\omega), \\ && \omega = e^{-t} [f'' u_x -f'
u_{xx}].
\end{eqnarray*}
\begin{equation}\label{3-dim-solv-6}
A_{3,6} : \quad J = \left(
{\begin{array}{*{20}c}
   1 & 0  \\
   0 & -1  \\
\end{array} } \right).
\end{equation}

\begin{eqnarray*}
A_{3,6}^{1}&: &X_1=\gen t,\quad X_2=\gen u,\quad
X_3=t\gen t+\frac{x}{3} \gen x-u\gen u,\\
&& F=x^{-6}\tilde F(x^{4}u_{x},x^{5}u_{xx}),\\
A_{3,6}^{2}&: &X_1=\gen x,\quad X_2=\gen u+\lambda t^{2/3}\gen x
,\quad
X_3=3t\gen t+x \gen x-u\gen u,\\
&& F=-\frac{2\lambda}{3}t^{-1/3}uu_x+t^{-4/3}\tilde
F(t^{2/3}u_{x},t
u_{xx}),\\
A_{3,6}^{3}&: &X_1=\gen u,\quad X_2=e^{2t}f(x)\gen u
,\quad
X_3=\gen t+u\gen u,\quad f'\ne 0,\\
&&
F=(2f-f''')(f')^{-1}u_x+e^{t}\tilde F(x,\omega),\quad
\omega=e^{-t}(f''u_x-f'u_{xx}),\\
A_{3,6}^{4}&: &X_1=\gen u,\quad X_2=e^{2f^{-1}x}h(t)\gen u
,\quad
X_3=f(t)\gen x+u\gen u,\quad fh\ne 0,\\
&& F = -[4 f^{-2} -\frac{1}{2} hh^{-1} f+ f^{-1} f' x] u_x
+e^{f^{-1}x} \tilde F(t, \omega),
\\
&&\omega=e^{-f^{-1}x}(2 u_x -f u_{xx}).
\end{eqnarray*}
\begin{equation}\label{3-dim-solv-7}
A_{3,7} : \quad J = \left(
{\begin{array}{*{20}c}
   1 & 0  \\
   0 & q  \\
\end{array} } \right),\quad 0<|q|<1.
\end{equation}

\begin{eqnarray*}
A_{3,7}^{1}&: & X_1=\gen t,\quad X_2=\gen x,\quad
X_3=t\gen t+\frac{x}{3} \gen x, \quad  q=1/3 \\
&& F=u_{x}^3\tilde F(u,u_{x}^{-2}u_{xx}),\\
A_{3,7}^{2}&: & X_1=\gen t,\quad X_2=\gen x,\quad
X_3=t\gen t+\frac{x}{3}\gen x+u\gen u,\quad q=1/3 \\
&& F=\tilde F(\omega_1,\omega_2),\quad \omega_1=u^{-2/3}u_x,\;
\omega_2=u^{-1/3}u_{xx},\\
A_{3,7}^{3}&: &X_1=\gen t,\quad X_2=\gen u,\quad
X_3=t\gen t+\frac{x}{3} \gen x+q u\gen u,\quad q\ne 0,\pm 1,\\
&& F=x^{3(q-1)}\tilde
F(\omega_1,\omega_2),\;\omega_1=x^{1-3q}u_x,\;
\omega_2=x^{2-3q}u_{xx},\\
A_{3,7}^{4}&: &X_1=\gen x,\quad X_2=\gen u+\lambda t^{(1-q)/3}\gen
x ,\quad X_3=3t\gen t+x \gen x+q u\gen u,\\
&& q\ne 0, \pm 1,\;\lambda\in\mathbb{R},\\
&&
F=\frac{\lambda}{3}(q-1)t^{-(q+2)/3}uu_x+\tilde F(\omega_1,\omega_2),\\
&&\omega_1=t^{-(q-1)/3}u_x,\quad\omega_2=t^{-(q-2)/3}u_{xx}.
\end{eqnarray*}

\begin{eqnarray*}
A_{3,7}^5 &:& X_1 = \partial_u, X_2 = e^{(1-q)t} f(x) \partial_u,
X_3 = \partial_t +u \partial_u, \quad f' \not =0, q\not =0, \pm 1, \\
&& F = [(1-q)f -f'''] (f')^{-1} u_x +e^t \tilde F(x, \omega), \\
&& \omega = e^{-t} [f''u_x -f' u_{xx}]; \\
A_{3,7}^6 &:& X_1 = \partial_u, \quad X_2 = e^{(1-q) f^{-1}(t)x} h(t)
\partial_u, \\
&& X_3 = f(t) \partial_x+u \partial_u, \quad f\cdot h \not =0, \quad
q\not=0,\;\pm 1,\\
&& F = -[(1-q)^2 f^2 +f^{-1} f' x-(1-q)^{-1} f h^{-1} h']u_x +e^{f^{-1}x}
\tilde F(t, \omega), \\
&& \omega = e^{-f^{-1}x} [(1-q) u_x -f u_{xx}].
\end{eqnarray*}

{\bf Remark:} The algebra $A_{3,7}$ has another realization
$$\curl{X_1=\gen t,
X_2=\gen x, X_3=t\gen t+\frac{1}{3}(x+b_0t)\gen x+u\gen u}$$
that is isomorphic to $A_{3,7}^2$ under the change of basis
$$X_1\to X_1+\frac{b_0}{2} X_2,\quad X_2 \to X_2, \quad X_3 \to
X_3.$$  Note that its equivalence to the latter is established by
the change of variables
$$\tilde t = t, \quad \tilde x =x-\frac{1}{2}b_0 t, \quad \tilde u =
u.$$ That is why, we have excluded it from the above list.

\begin{equation}\label{3-dim-solv-8}
A_{3,8} : \quad J = \left(
{\begin{array}{*{20}c}
   0 & -1  \\
   1 & 0  \\
\end{array} } \right).
\end{equation}

\begin{eqnarray*}
A_{3,8}^{1}&: &X_1=\gen x,\quad X_2=\alpha(t)\gen x+ \gen u,\\
&& X_3=-\frac{1}{\dot{\alpha}}(1+\alpha^2)\gen t-\alpha x
\gen x+(\alpha u-x)\gen u,\\
&& F=-\dot{\alpha}u u_x+(1+\alpha^2)^{-2}\tilde F(\omega_1,\omega_2),\\
&&\omega_1=(1+\alpha^2)u_x-\alpha,\;\omega_2=(1+\alpha^2)^{3/2}
u_{xx},
\end{eqnarray*}
where $\alpha(t),$ $\dot\alpha\ne 0$ satisfies
\begin{equation}\label{odex}
(1+\alpha^2)\ddot{\alpha}+\alpha{\dot{\alpha}}^2=0.
\end{equation}

\begin{equation}\label{3-dim-solv-9}
A_{3,9} : \quad J = \left(
{\begin{array}{*{20}c}
   q & -1  \\
   1 & q  \\
\end{array} } \right),\quad q>0.
\end{equation}

\begin{eqnarray*}
A_{3,9}^{1}&: &X_1=\gen x,\quad X_2=\alpha(t)\gen x+ \gen u,\\
&& X_3=-\frac{1}{\dot{\alpha}}(1+\alpha^2)\gen t+(q-\alpha)x
\gen x+[(q+\alpha)u-x]\gen u,\\
&& F=-\dot{\alpha}u u_x+\exp{\curl{2q \arctan \alpha
}}(1+\alpha^2)^{-2}\tilde F(\omega_1,\omega_2),\\
&&\omega_1=(1+\alpha^2)u_x-\alpha,\;\omega_2=(1+\alpha^2)^{3/2}
\exp{\curl{-q\arctan{\alpha}}}u_{xx},
\end{eqnarray*}
where $\alpha(t),$ $\dot\alpha\ne 0$ satisfies
\begin{equation}\label{ode}
(1+\alpha^2)\ddot{\alpha}+(\alpha-3q){\dot{\alpha}}^2=0.
\end{equation}

{\bf Remark:} $\alpha(t)$ can be obtained implicitly by
quadratures as
$$
\int_{}^\alpha  {\exp ( - 3q\arctan \xi )(1 + \xi ^2 )^{1/2}d\xi  = c_1
t + c_0 }.
$$

\begin{thm}
There are thirty-eight inequivalent three-dimensional solvable symmetry
algebras admitted by equation \eqref{main}.
\end{thm}
\section{Equations with four-dimensional solvable algebras}
For $\dim L=4$, we proceed exactly in the same manner as above.
We start from the already standardized three-dimensional algebras,
and add a further linearly independent element $X_4$, and require
that they form a Lie algebra.

{\bf Decomposable algebras:} The list of decomposable
four-dimensional Lie algebras consists of the twelve algebras:\ $4 A_1 =
A_{3,1}\oplus A_1,\ A_{2,2}\oplus 2A_1= A_{3,2}\oplus A_1,\ 2
A_{2,2}=A_{2,2}\oplus A_{2,2},\ A_{3,i}\oplus A_1\ (i = 3,4,
\ldots 9)$. We preserve the notations of the previous section.

There are four inequivalent realizations of the algebra $2A_{2,2}$
which are invariance algebras of PDEs of the form (\ref{main}). We
give these realizations  together with the corresponding invariant
equations.

\begin{eqnarray*}
2A_{2,2}^{1}&: &\quad X_1=-t\gen t-\frac{x}{3}\gen x,\quad
X_2=\gen t,\quad
X_3=\gen u,\quad X_4=e^u \gen u,\\
&&F=u_x^3-3u_xu_{xx}+x^{-2}u_x \tilde F(\omega),\quad
\omega=x({u^{-1}_x}u_{xx}-u_{x}),\\
2A_{2,2}^{2}&: &\quad X_1=-3t\gen t-x\gen x,\quad X_2=\gen x,\quad
X_3=-u\gen u+\lambda t^{1/3}\gen x,\quad X_4=\gen u,\\
&&F=\frac{\lambda}{3t}\omega_1 \ln|\omega_1|
+\frac{\omega_1}{t}\tilde F(\omega),\quad
\omega_1=t^{1/3}u_x,\quad \omega=t^{1/3}{u^{-1}_x}u_{xx},\\
2A_{2,2}^{3}&: &\quad X_1=\gen x-u\gen u,\quad X_2=\gen u,\quad
X_3=\frac{1}{\lambda} \gen t,\quad X_4=\exp{(\lambda t)}\gen x,\\
&&F=-{\lambda}{x}u_x-\lambda u_x \ln|u_x|+u_x\tilde
F(\omega),\quad
\omega={u^{-1}_x}u_{xx},\\
2A_{2,2}^4 &:&  X_1 = \partial_x-u \partial_u, \quad X_2 = \partial_u, \quad
X_3 = \lambda \partial_t, \quad  X_4 = e^{\lambda^{-1} t-x} \partial_u,
\quad
\lambda \not =0,\\
&& F = (1+\lambda^{-1}) u_x +e^{-x} \tilde F(\omega), \quad \omega = e^x
(u_x +u_{xx}),\\
2A_{2,2}^5 &:& X_1 = \partial_t-u \partial_u, \quad X_2 = \partial_u, \quad
X_3 = \beta (\partial_x+\gamma \partial_t) -\partial_t, \\
&& X_4 = e^{\gamma x-t} \partial_u, \quad  \beta  \gamma \not =0, \\
&& F = e^{(\gamma -\beta^{-1})x-t} \tilde F(\omega) -\gamma^{-1}
(1+\gamma^3) u_x, \\
&& \omega = e^{t+(\beta^{-1} -\gamma)x} (\gamma u_x -u_{xx}).
\end{eqnarray*}

\noindent {\bf Equations invariant under the algebra} $A_{2,2} \oplus 2 A_1
=
A_{3.2} \oplus A_1$:

\begin{eqnarray*}
A^6_{3,2} \oplus \{X_4 \} & : & X_1 = \partial_x-u \partial_u, \quad
X_2 = \partial_u, \quad X_3 = \partial_t, \quad X_4 = e^{-x} \partial_u,\\
&& F = -u_x +e^{-x} \tilde F(\omega), \quad \omega = e^x (u_x +u_{xx}), \\
A^6_{3,2} \oplus \{ X_4 \} & : & X_1 = \partial_x-u \partial_u, \quad
X_2 = \partial_u, \quad X_3 = \partial_t, \quad X_4 = \partial_x, \\
&& F = u_x \tilde F(\omega), \quad  \omega = u_{xx} u^{-1}_x,\\
\end{eqnarray*}
\begin{eqnarray*}
A^7_{3,2}  (f = e^{\lambda x}, \ \lambda \not =0) \oplus \{ X_4 \}& : &
X_1 = \partial_t -u \partial_u, \quad X_2 = \partial_u, \\
&& X_3 = e^{\lambda x-t} \partial_u, \quad  X_4 = \partial_x +\lambda
\partial_t, \quad \lambda \not =0, \\
&& F = -(\lambda^3 +1) \lambda^{-1} u_x +e^{-t+\lambda x} \tilde
F(\omega),\\  &&\omega = e^{t-\lambda x} (\lambda u_x -u_{xx}). \\
\end{eqnarray*}

\begin{eqnarray*}
A^1_{3.3} \oplus\{ X_4 \} &:& X_1 = \partial_t, \quad X_2 = \partial_u,\quad
X_3 = t\partial_u+\lambda \partial_x, \\
&&  X_4 = \partial_t+\lambda^{-1} x  \partial_u+\beta \partial_x, \quad
\lambda >0, \quad
\beta \in \R,\\
&& F = \lambda^{-1} x-\beta u_x + \tilde F(u_{xx}), \\
A^3_{3,5} \oplus \{ X_4 \} & : & X_1 = \partial_u, \quad X_2 = \partial_x,
\quad
X_3 = x\partial_u+\lambda \partial_t, \quad X_4 = \partial_t+
\beta (\partial_x+\lambda^{-1} t \partial_u), \\
&& \lambda \not =0, \quad  \beta \in \R,\quad
F = \beta (\lambda^{-1}t-3 u_x) + \tilde F(u_{xx}),\\
\end{eqnarray*}
\begin{eqnarray*}
A^6_{3,5}  (f = \lambda^{-1} x, \ \lambda \not =0) \oplus \{ X_4 \} &:&
X_1 = \partial_u, \quad X_2 = (\lambda^{-1} x-t) \partial_u, \\
&& X_3 =  \partial_t, \quad  X_4 = \partial_t +\lambda \partial_x, \quad
\lambda \not =0, \\
&& F = -u_x + \tilde F(u_{xx}), \\
\end{eqnarray*}
\begin{eqnarray*}
A^9_{3,5} \oplus \{ X_4 \} &:& X_1 = -x^{-1} \partial_u, \quad X_2 =
\partial_u, \quad
X_3 = \partial_x -x^{-1} u \partial_u, \quad X_4 = \partial_t, \\
&& F = 3 x^{-1} u_{xx} +x^{-1} \tilde F(\omega), \quad  \omega = 2 u_{x} +x
u_{xx}.
\end{eqnarray*}

\begin{eqnarray*}
A^1_{3,4} \oplus \{ X_4 \} &:&  X_1 = \partial_u, \ X_2 = \partial_t, \\
&& X_3 = t \partial_t +\frac{1}{3} x \partial_x +(u+t) \partial_u, \ \ X_4 =
x^3 \partial_u, \\
&& F = 3 \ln x -2 x^{-2} u_x +\tilde F(\omega), \ \omega = x^{-1} u_{xx} -2
x^{-2} u_x, \\
A^5 _{3,4} \ (f = \lambda x, \ \lambda \not =0) \oplus \{ X_4 \} &:&  X_1 =
\partial_u, \\
&& X_2 = (-t +\lambda x) \partial_u, \quad X_3 = \partial_t +u \partial_u,
\\
&& X_4 = \partial_x+\lambda \partial_t, \quad \lambda \not =0, \\
&& F = -\lambda^{-1} u_x +e^{t-\lambda x} \tilde F(\omega), \ \omega = e^{-t
+\lambda x} u_{xx}, \\
A^6_{3,4} \oplus \{ X_4 \} &:& X_1 = \partial_u, \quad X_2 = -x \partial_u,
\\
&& X_3 = \partial_x +u \partial_u, \quad X_4 = \partial_t, \\
&& F = e^x \tilde F(\omega), \quad \omega = e^{-x} u_{xx}.
\end{eqnarray*}

\begin{eqnarray*}
A^1_{3,5} \oplus \{ X_4 \} &:& X_1 = \partial_t, \quad X_2 = \partial_u, \\
&& X_3 = t \partial_t +\frac{x}{3} \partial_x +u \partial_u, \quad X_4 = x^3
\partial_u, \\
&& F = -2 x^{-2} u_x +\tilde F(\omega), \quad   \omega = x^{-1} u_{xx} -2
x^{-2} u_x.
\end{eqnarray*}

\begin{eqnarray*}
A^1_{3,6} \oplus \{ X_4 \} &:& X_1 =\partial_t, \quad X_2 = \partial_u, \\
&& X_3 = t \partial_t+\frac{x}{3} \partial_x-u \partial_u, \quad   X_4 =
x^{-3} \partial_u, \\
&& F = -20 x^{-2} u_x +x^{-6} \tilde F(\omega), \quad \omega = 4 x^4 u_x
-x^5 u_{xx},\\
A^3_{3,6} (f = e^{-2 \beta^{-1}x}, \ \beta \not =0) \oplus \{ X_4 \} &:& X_1
= \partial_u, \quad X_2 = e^{2(t-\beta^{-1}x)} \partial_u, \\
&& X_3 = \partial_t+u \partial_u, \quad X_4 = \partial_t+\beta \partial_x,
\quad \beta \not =0, \\
&& F = -(\beta +4 \beta^{-2}) u_x +e^{t-\beta^{-1}x} \tilde F(\omega), \\
&& \omega = e^{-t+\beta^{-1} x}(2 u_x +\beta u_{xx}), \\
A^4_{3,6} (f=h=1) \oplus \{ X_4 \} &:& X_1 = \partial_u, \quad X_2 = e^{2x}
\partial_u, \\
&& X_3 = \partial_x+u \partial_u, \quad X_4 = \partial_t, \\
&& F = -4 u_x +e^x \tilde F(\omega), \quad  \omega = e^{-x} (2 u_x -u_{xx}).
\end{eqnarray*}

\begin{eqnarray*}
A^1_{3,7} \oplus \{ X_4 \} &:& X_1 = \partial_t, \quad X_2 = \partial_x, \\
&& X_3 = t \partial_t +\frac{1}{3} x \partial_x, \quad X_4 = u\partial_u, \\
&& F = u^{-2}u^3_x \tilde F(\omega), \quad \omega = u_x^{-2}u u_{xx}, \\
A^3_{3,7} \oplus \{ X_4 \} &:& X_1 = \partial_t, \quad X_2 = \partial_u, \\
&& X_3 = t \partial_t +\frac{x}{3} \partial_x +q u \partial_u, \quad X_4 =
x^{3q} \partial_u,\\
&&   q\not =0, \pm 1, \\
&& F = -(3q-1) (3q-2) x^{-2} u_x +x^{3(q-1)} \tilde F(\omega), \\
&& \omega = x^{1-3q} [(3q-1) u_x -x u_{xx}], \\
A^5_{3,7} (f = e^{-1(1-q)\beta^{-1}x} , \beta \not =0) \oplus \{ X_4
\} &:& X_1 = \partial_u, \quad X_2 = e^{(1-q)(t-\beta^{-1}x)} \partial_u,
\quad X_3=\partial_t+u \partial_u, \\
&& X_4 = \partial_t +\beta \partial_x, \quad \beta \not =0, \ q\not =0, \pm
1, \\
&& F = -[\beta +(1-q)^2 \beta^{-2}] u_x +e^{t-\beta^{-1}x} \tilde F(\omega),
\\
&& \omega = e^{-t+\beta^{-1}x} [(1-q) u_x +\beta u_{xx}], \\
A^6_{3,7} (f =h=1) \oplus \{ X_4 \} &:& X_1 = \partial_u, \quad X_2 =
e^{(1-q)x} \partial_u, \\ && X_3 = \partial_x+u \partial_u, \quad X_4 =
\partial_t, \quad q\not =0, \pm 1,\\
&& F = -(1-q)^2 u_x +e^x \tilde F(\omega), \\
&& \omega = e^{-x} [(1-q)u_x -u_{xx}].
\end{eqnarray*}

\noindent{\bf Remark:} The $A^1_{3,7}\oplus A_1$ invariant
equation is
\begin{equation}\label{inv}
u_t=u_{xxx} + \frac{{u_x^3 }} {{u^2 }}\tilde F(\omega),\quad
\omega=\frac{{uu_{xx} }} {{u_x^2 }}.
\end{equation}
If, in particular, $\tilde{F}=c\omega^2$, $c=\text{const.}$,
namely
\begin{equation}\label{F0}
F=cu_x^{-1}u_{xx}^2
\end{equation}
the symmetry algebra is further  extended by $X_5={\gen u}$ to a
five-dimensional one.

{\bf Nondecomposable algebras:}

The set of inequivalent abstract four-dimensional Lie algebras
contains ten real non-de\-com\-pos\-able Lie algebras $A_{4,i} =
\curl{X_1$, $X_2, X_3$, $X_4}$ $(i=1,\ldots, 10)$ \cite{Zhdanov99,
Basarab-Horwath01}. They are all solvable and therefore can be written
as semidirect sums of a one-dimensional Lie algebra $\curl{X_4}$
and a three-dimensional ideal $N=\curl{X_1,X_2,X_3}$. For
$A_{4,i}$ $(i=1,\ldots, 6)$, $N$ is abelian, for $A_{4,7},
A_{4,8}, A_{4,9}$ it is of type $A_{3,3}$ (nilpotent), and for
$A_{4,10}$ it is of the type $A_{3,5}$. The non-zero commutation
relations read as
\begin{eqnarray*}
A_{4,1}&:& [X_2, X_4] = X_1, \quad [X_3, X_4] = X_2; \\
A_{4,2}&:& [X_1, X_4] = qX_1,\quad [X_2, X_4] = X_2, \\
&& [X_3, X_4] = X_2 +X_3, \quad q\not =0;\\
A_{4,3}&:& [X_1, X_4] = X_1,\quad [X_3, X_4] = X_2; \\
A_{4,4}&:& [X_1, X_4] = X_1,\quad [X_2, X_4] = X_1+X_2, \\
&& [X_3, X_4] = X_2 +X_3;\\
A_{4,5}&:& [X_1, X_4] = X_1,\quad [X_2, X_4] =q X_2, \\
&& [X_3, X_4] = p X_3,\quad -1\le p \le q \le 1, \quad p q \not
=0;\\
A_{4,6}&:& [X_1, X_4] = qX_1,\quad [X_2, X_4] = pX_2-X_3, \\
&& [X_3, X_4] = X_2 +p X_3,\quad q\not =0,\quad p \ge 0;\\
A_{4,7}&:& [X_2, X_3] = X_1,\quad [X_1, X_4] = 2X_1, \\
&& [X_2, X_4] = X_2,\quad [X_3, X_4] = X_2 +X_3;\\
A_{4,8}&:& [X_2, X_3] = X_1,\quad [X_1, X_4] = (1+q) X_1, \\
&& [X_2, X_4] = X_2 ,\quad [X_3, X_4] = q X_3, \quad |q|
\le 1;\\
A_{4,9}&:& [X_2, X_3] = X_1, \quad [X_1, X_4] = 2q X_1, \\
&& [X_2, X_4] = q X_2-X_3,\quad [X_3, X_4] = X_2+ q X_3,
\quad q \ge 0; \\
A_{4,10}&:& [X_1, X_3] = X_1,\quad [X_2, X_3] = X_2, \\
&& [X_1, X_4] = -X_2,\quad [X_2, X_4] = X_1.
\end{eqnarray*}

In order to obtain realizations of solvable four-dimensional symmetry
algebras of PDEs that belong to the class \eqref{main}, we add
$X_4$ in the generic form \eqref{gvf} to the already constructed
three-dimensional symmetry algebras and impose the above
commutation relations. Once the algebra is found we insert
$X_4$ into equation \eqref{deq} and solve it for the
function $F$. The form of $F$ which is invariant under a
three-dimensional algebra is further restricted.

\begin{eqnarray*}
A_{4,1}^{1}&: &\quad X_1=\gen u,\quad X_2=\gen x,\quad
X_3=\gen t,\quad X_4=t \gen x+x\gen u,\\
&&F=-\frac{1}{2}u_x^2+\tilde F(u_{xx}),\\
A^2_{4,1} &:& X_1 = \partial_u, \quad X_2 = x \partial_u,
\quad X_3 = \partial_t, \quad X_4 = \partial_x+tx \partial_u, \\
&& F = \frac{1}{2} x^2 +\tilde F(u_{xx}),\\
A_{4,2}^{1}&: &\quad X_1=\gen t,\quad X_2=\gen u,\quad
X_3=\gen x,\quad X_4=3t \gen t+x\gen x+(x+u)\gen u,\\
&&F=u_{xx}^2\tilde F(e^{u_x}u_{xx}),\\
A_{4,2}^{2}&: &\quad X_1=\gen x,\quad X_2=\gen u,\quad
X_3=\gen t,\quad X_4=t \gen t+\frac{x}{3}\gen x+(t+u)\gen u,\\
&&F=\frac{3}{2}\ln |u_x|+\tilde F(\omega),\quad
\omega=\frac{u_{xx}^2}{u_x},\\
A^3_{4,2} &:& X_1 = \partial_t, \quad X_2 =  \partial_u, \quad X_3 = -3
q^{-1} \ln x \partial_u, \\
&& X_4 = qt \partial_t+\frac{1}{3} qx  \partial_x+ u \partial_u, \quad q\not
=0, \\
&& F = -2 x^{-2}u_x +x^{3(q^{-1}-1)} \tilde F(\omega ), \quad \omega =
x^{1-3q^{-1}} u_x +x^{2-3q^{-1}} u_{xx},\\
A^4_{4,2} &:& X_1 = x^{3(1-q)}\partial_u, \quad X_2 = \partial_u, \quad X_3
= \partial_t, \\
&& X_4 = t\partial_t +\frac{1}{3} x\partial_x + (u+t) \partial_u, \quad
q\not =0,1, \\
&& F = -(2-3q)(1-3q) x^{-2} u_x +3 \ln x+ \tilde F(\omega), \quad \omega =
(2-3q)x^{-2} u_x -x^{-1} u_{xx},\\
A_{4,3}^{1}&: &\quad X_1=\gen u,\quad X_2=\gen x,\quad
X_3=\gen t,\quad X_4=t \gen x+u\gen u,\\
&&F=-u_x\ln |u_x|+u_x\tilde F(\frac{u_{xx}}{u_x}),\\
A^2_{4,3} &:& X_1 = \partial_t, \quad X_2 =  \partial_u, \quad X_3 = -3 \ln
x \partial_u, \\
&& X_4 = t \partial_t+\frac{1}{3} x  \partial_x, \\
&& F = -2 x^{-2}u_x +x^{-3} \tilde F(\omega ), \quad \omega = x u_x +x^{2}
u_{xx},\\
A^3_{4,3} &:& X_1 = \partial_u, \quad X_2 = e^x\partial_u, \quad X_3 =
\partial_t, \\
&& X_4 = \partial_x +(u+t e^x) \partial_u,  \\
&& F = -u_x +xe^x + e^x \tilde F(\omega), \quad \omega = e^{-x}( u_x
-u_{xx}),\\
A^1_{4,4} &:& X_1 = \partial_u, \ X_2 = -3\ln \, x \partial_u, \ X_3 =
\partial_t, \\
&& X_4 = t \partial_t+\frac{1}{3} x  \partial_x+(u-3 t \ln \, x) \partial_u,
\\
&& F = -2 x^{-2}u_x -\frac{9}{2} \ln^2 x + \tilde F(\omega ), \quad \omega =
x^{-2} u_x +x^{-1} u_{xx},\\
\end{eqnarray*}
\begin{eqnarray*}
A_{4,5}^{1}&: &\quad X_1=\gen t,\quad X_2=\gen x,\quad
X_3=\gen u,\quad X_4=t \gen t+\frac{x}{3}\gen x+ku\gen u,\quad k\ne
0,\frac{1}{3},\\
&&F=u_x^{3(1-k)/(1-3k)}\tilde F(\omega),\quad
\omega=u_x^{(3k-2)/(1-3k)}u_{xx},\\
A^2_{4,5} &:& X_1 = \partial_t, \quad X_2 =  \partial_x ,\quad X_3 =
\partial_u, \\
&& X_4 = t \partial_t+\frac{x}{3}   \partial_x+\frac{u}{3} \partial_u, \\
&& F =  u^{2}_{xx} \tilde F(u_{xx} ) , \\
A^3_{4,5}&:& X_1 = \partial_t, \quad X_2 = u, \ X_3 = x^{3(q-p)} \partial_u,
\\
&& X_4 = t\partial_t +\frac{1}{3} x\partial_x +qu \partial_u, \quad  q\not
=p, \quad q \cdot p \not =0, \\
&& F = -[3 (q-p) -1] [3(q-p) -2] x^{-2} u_x+x^{3(q-1)} \tilde F(\omega), \\
&& \omega = [3(q-p) -1] x^{1-3q} u_x -x^{2-3q} u_{xx},\\
A_{4,7}^{1}&: &\quad X_1=\gen u,\quad X_2=\gen x,\quad
X_3=x\gen u-\frac{1}{3}\ln t\gen x,\quad X_4=3t \gen t+x\gen x+2u\gen u,\\
&&F=\frac{1}{6t}u_x^2+t^{-1/3}\tilde F(u_{xx}),\\
A^2_{4,7} &:& X_1 = \partial_u, \quad \ X_2 = x\partial_u +b \partial_x,
\quad X_3 = -\partial_x,\\
&& X_4 = -b^2 (b')^{-1} \partial_t+(1-b)x \partial_x+(2u-\frac{1}{2} x^2)
\partial_u, \quad b=b(t), \ b' \not =0, \\
&& b^2 b'' +(b-3)(b')^2 =0, \\
&& F = -\frac{1}{2} b' u^2_x +b^{-3} e^{-b^{-1}} \tilde F(\omega),
\quad
\omega = b^{2} u_{xx} -b;\\
A^3_{4,7} &:& X_1 = \partial_u, \quad X_2 = (\lambda x^3 -t) \partial_u,
\quad X_3 = \partial_t, \\
&& X_4 = t \partial_t +\frac{1}{3} x \partial_x +\left(2u -\frac{1}{2} t^2
+\lambda t x^3\right) \partial_u, \quad \lambda \not =0, \\
&& F = -\frac{1}{3} \lambda^{-1} (1+6 \lambda) x^{-2} u_x +3 \lambda x^3 \ln
x+x^3 \tilde F (\omega), \quad
\omega = 2 x^{-5} u_x -x^{-4} u_{xx},\\
A_{4,7}^{4}&: &\quad X_1=\gen u,\quad X_2=(t-x)\gen u,\quad
X_3=\gen x,\\
&&X_4=3t \gen t+(x+2t)\gen x+(xt-\frac{x^2}{2}+2u)\gen u,\\
&&F=-u_x+t^{-1/3}\tilde F(\omega)+\frac{t}{4},\quad
\omega=u_{xx}+\frac{1}{3}\ln
t,\\
A_{4,7}^{5}&: &\quad X_1=\gen u,\quad X_2=-x\gen u,\quad
X_3=\gen x,\\
&&X_4=3t \gen t+x\gen x+(2u-\frac{x^2}{2})\gen u,\\
&&F=t^{-1/3}\tilde F(\omega),\quad \omega=u_{xx}+\frac{1}{3}\ln t,\\
\end{eqnarray*}
\begin{eqnarray*}
A^6_{4,7}&:& X_1= -x^{-1} \partial_u, \quad X_2 = \partial_u, \quad X_3 =
\partial_x-x^{-1} u \partial_u, \\
&& X_4 = 3t \partial_t +x \partial_x+\left(u +\frac{1}{2} x\right)
\partial_u, \\
&& F = 3 x^{-1} u_{xx} +x^{-1} t^{-\frac{1}{3}} \tilde F(\omega),
\quad \omega = 2 u_x +xu_{xx} -\frac{1}{3} \ln t,\\
A_{4,8}^{1} &:
&\quad X_1=\gen x,\quad X_2=\gen t,\quad X_3=t\gen x+\gen u,\quad
X_4=t \gen t+\frac{x}{3}\gen x-\frac{2}{3}u\gen u,
\\
&&F=-uu_x+u_x^{5/3}\tilde F(u_x^{-4/3}u_{xx}),\\
A_{4,8}^{2} &: &\quad X_1=\gen u,\quad X_2=\gen t,\quad
X_3=t\gen u+\lambda\gen x,\quad X_4=t \gen t+\frac{x}{3}\gen x+
\frac{4}{3}u\gen u,\\
&&F=\frac{x}{\lambda}+u_x^{1/3}\tilde F(\omega),\quad
\omega=u_x^{-2/3}u_{xx},
\quad \lambda>0\\
A_{4,8}^{3} &: &\quad X_1=\gen u,\quad X_2=\gen x,\quad X_3=x\gen
u+\lambda t^{(1-q)/3}\gen x, \\
&&\quad X_4=3t \gen t+x\gen x+(1+q)u\gen u, \quad q\in\R\\
&&F=\frac{\lambda(q-1)}{6}t^{-(2+q)/3}u_x^{2}+t^{(q-2)/3}\tilde
F(\omega),
\quad \omega=t^{(1-q)/3}u_{xx},\quad \lambda\ne 0,\quad |q|\ne 1,\\
A^4_{4,8} &:& X_1 = \partial_u, \quad X_2 = \partial_x, \quad X_3 =
x\partial_u+\lambda \partial_t, \\
&& X_4 = 3t \partial_t+x \partial_x+4u \partial_u, \quad \lambda \not =0, \\
&& F = (t-3\lambda u_x)^{\frac{1}{3}} \tilde F(\omega), \quad \omega=
u^3_{xx} (t-3 \lambda u_x)^2,\\
A^5_{4,8} &:& X_1 = \partial_u,\quad X_2 = (\lambda x^3 -t)\partial_u, \quad
X_3 = \partial_t, \\
&& X_4 = qt \partial_t +\frac{1}{3} qx \partial_x+(1+q)u \partial_u, \quad
\lambda \cdot q\not =0, \\
&& F = -\frac{1}{3} \lambda^{-1} (1+6 \lambda) x^{-2} u_x +x^{3q^{-1}}
\tilde F(\omega), \quad \omega = 2 x^{-(2+3q^{-1})}u_x-x^{-(1+3q^{-1})}
u_{xx}, \\
A_{4,8}^{6} &: &\quad X_1=\gen u,\quad X_2=(t-x)\gen u,\quad
X_3=\gen x, \\
&&\quad X_4=3qt \gen t+q(x+2t)\gen x+(1+q)u\gen u,\\
&&F=-u_x+t^{\frac{1}{3}(1-2q)q^{-1}}\tilde F(\omega), \quad
\omega=t^{\frac{1}{3}(q-1)q^{-1}}u_{xx},\\
A^7_{4,8} &:& X_1 = -x^{-1} \partial_u, \quad X_2 = \partial_u, \quad X_3 =
\partial_x-x^{-1} u \partial_u, \\
&& X_4 = 3qt +qx \partial_x+u \partial_u, \quad q\not =0, \\
&& F = 3x^{-1} u_{xx} +x^{-1} t^{\frac{1}{3}(q-1) q^{-1}} \tilde
F(\omega), \quad \omega = t^{\frac{1}{3}(q-1) q^{-1}} (2 u_x +x
u_{xx}),\\
A_{4,8}^{8} &: &\quad X_1=\gen u,\quad X_2=-x\gen u,\quad
X_3=x\gen x, \quad
X_4=3qt \gen t+qx\gen x+(1+q)u\gen u,\\
&&F=t^{\frac{1}{3}(1-2q)q^{-1}}\tilde F(\omega), \quad
\omega=t^{\frac{1}{3}(q-1)q^{-1}}u_{xx}.
\end{eqnarray*}

{\bf Remarks:}

\begin{itemize}
\item There exists a realization of $A_{4,6}$.

\begin{eqnarray*}
A^1_{4,6} &:& X_1 = \partial_t, \quad X_2 =  \tan \psi \partial_u ,\quad X_3
=  \partial_u, \\
&& X_4 = 2t \partial_t+\frac{2}{3} x  \partial_x+[p+\tan \psi] u \partial_u,
\\
&& \psi = \frac{3}{2} \ln x, \quad  \ p \in \R.
\end{eqnarray*}
However there are no equations that can be invariant under this algebra.

\item The algebra $A_{4,8}^1$ is isomorphic to the KdV algebra which
is the semidirect sum of the  nilradical (maximal nilpotent ideal)
$\nil(2)=\curl{X_1, X_2, X_3}$ and the dilation $\Di=\curl{X_4}$.
\end{itemize}
\[
\begin{aligned}
  A_{4,9}^1 \quad :\quad X_1  &  = \partial _u ,\quad X_2  = \partial _x
,\quad
  X_3  =    \alpha (t)\partial _x + x\partial _u , \\
  X_4  &  =  - \frac{{(1 + \alpha ^2 )}}
{{\dot \alpha }}\partial _t  + (q - \alpha )x\partial _x  + (2qu -
\frac{{x^2 }}
{2})\partial _u ,\quad q\in\R \\
  F &  =  - \frac{1}
{2}\dot \alpha u_x^2  +  (1 + \alpha ^2 )^{ - 3/2}\exp(q
\arctan\alpha) \tilde F(\omega ),\quad \omega  =
(1 + \alpha ^2 )u_{xx}  - \alpha , \\
   & (1 + \alpha ^2 )\ddot \alpha  + (\alpha  - 3q)\dot \alpha ^2  = 0. \\
\end{aligned}
\]
The function $\alpha(t),$ $\dot\alpha\ne 0$ is a solution of the
\ode{} \eqref{ode}.


\[
\begin{aligned}
A^1_{4,10} \quad :\quad X_1 & = \partial_u, \quad X_2= -\tan x
\partial_u,
\quad X_3 =\partial_t+u \partial_u, \\
X_4 & = \beta \partial_t+ \partial_x+u \tan x \partial_u,\quad
\beta \in \R,\\
F & = -2 u_x -3 \tan x u_{xx} +e^{t-\beta x} \sec x\tilde F (\omega), \\
\omega &= e^{\beta x-t} (\cos x u_{xx} -2  \sin x u_x).
\end{aligned}
\]

We sum up the above results as a theorem.

\begin{thm}
There exist fifty-two inequivalent four-dimensional symmetry
algebras admitted by equation \eqref{main}. The explicit forms of
those algebras as well as the associated invariant equations are
given above.
\end{thm}

\section{Discussion and Conclusions}
This paper provides a symmetry classification of the KdV type
equations involving an arbitrary function of five arguments. We
find that the equivalence classes of invariant equations involve
an arbitrary function of four, three, two variables and one
variable as soon as the symmetry algebra is one-, two-, three-,
and four-dimensional, respectively. In particular, we
studied symmetries of the  most general third-order linear
evolution equation. What came out from this, to our surprise, is
that the  symmetry group allowed is four-dimensional at most,
while there are nonlinear equations with symmetry algebras greater
than four. This result is in contrast to the second-order
evolution equations. It is exactly the linear heat equation that
allows for the maximal symmetry algebra.

To complete the classification list, it only remains to obtain the
inequivalent equations invariant under solvable algebras of the
dimension $\dim L\geq 5$. But this would require to go
through a large number of isomorphism classes. To give an idea of
the complexity of this task let us recall that there are sixty-six
classes of nonisomorphic real, solvable Lie algebras of dimension
five. For dimension six, there exist ninety-nine classes of them
with a nilpotent element. We plan to devote a separate article to
study equations admitting higher-dimensional symmetry algebras.

Whenever  $F$ is an arbitrary function of its arguments, the
symmetry algebras given in the paper are maximal. In particular,
if we impose the requirement that function $F$ is be independent of
$u_{xx}$ then we find that $\phi=R(t)u+S(x,t)$ in \eqref{gvf}. In this
case, invariance under four-dimensional algebras will force $F$
to depend on an arbitrary constant rather than on an arbitrary
function. Then, they may admit symmetry groups of the dimension higher
than four. We have analyzed this restricted class of equations and obtained
that
the only equation whose symmetry algebra is higher than four is
the one corresponding to the realization $A_{4,1}^{1}$ for $\tilde
F=\text{const.}$ On the other hand, for the specific choices of
$\tilde F$ involving one variable, the equations with
four-dimensional symmetry algebras may be invariant under larger
symmetry groups. For instance, the particular case of the equation
invariant under $A_{4,1}^{1}$ obtained by setting $\tilde
F=cu_{xx}^{4/3},$ $c=\text{const.}$ admits an additional symmetry
group generated by the dilation operator $X_5=3t\gen t+x\gen x-u\gen u$.

We only presented representative lists of equivalence classes of
invariant equations. All other invariant equations can be
recovered from these lists by applying the point transformations
\eqref{equ}. In other words, an equation in the class \eqref{main}
will have a symmetry group with dimension satisfying $\dim L\leq
6$ if and only if it can be transformed to one in the (canonical)
equations from the list.

As we mentioned, our classification is performed within point
transformations of coordinates. Two equations are equivalent
if one can be obtained from the other by a change of variables. On
the other hand, consider a special case of \eqref{F0} for $c=-3/4$
\cite{Sokolov84}
$$u_t=u_{xxx}-\frac{3}{4}\frac{u_{xx}^2}{u_x}$$
which additionally allows a symmetry group generated by
$\curl{\gen u}.$ Though this equation is equivalent to the
third-order linear equation $v_t=v_{xxx}$ under the (no-point)
transformation $v=\sqrt{u_x}$, we treat them as inequivalent.

To give a reader an insight into possible applications
of the results of this article, we consider a
subclass of equations \eqref{main}
\begin{equation}\label{subclass}
u_t=u_{xxx}+uu_x+f(t)u
\end{equation}
which arises in several physical applications such as propagation
of waves in shallow water of variable depth.

When $f(t)$ is arbitrary, \eqref{subclass} admits a
two-dimensional abelian  symmetry algebra generated by
\begin{equation}
X_1=\gen x,\quad X_2=\xi(t)\gen x-\dot\xi(t)\gen u,\quad \xi=\int
\exp\curl{\int f(t)\;dt}\; dt.
\end{equation}
By the change of dependent variable $\tilde u=u/\dot\xi$, the
generators are transformed to the realization $A_{2,1}^3$ with
$\alpha=-\xi.$ The corresponding invariant equation takes the form
    \[\tilde{u}_t=\tilde{u}_{xxx}+\dot{\xi}\tilde{u}\tilde{u}_x
\]
which is a particular case of \eqref{vcKdV}.

For the special case $f(t)=a t^k,\;(a\ne 0)$ the algebra is
larger and we have the following possibilities for the algebra to
be either three- or four-dimensional.

1.) $(a,k)=(a,-1),$  $a\ne -1$: The equation admits the
three-dimensional indecomposable solvable symmetry algebra spanned
by
\begin{equation}
X_1=\gen x,\quad X_2=t^{1+a}\gen x-(1+a)t^{a}\gen u,\quad
X_3=t\gen t+\frac{x}{3}\gen x-\frac{2}{3}u\gen u
\end{equation}
with non-zero commutation relations
$$[X_1,X_3]=-\frac{1}{3}X_1,\quad [X_2,X_3]=-\frac{3a+2}{3}X_2.$$

For $a=-1/3$, the algebra is isomorphic, up to scaling of basis
elements, to $A_{3,5}$, for $-1<a<-\frac{1}{3}$, to $A_{3,7}$.

For $a=-2/3$ it is isomorphic to the decomposable solvable algebra
$A_{3,2}$ and a suitable basis is
$$X_1=\gen x, \quad X_2=t^{1/3}\gen x-\frac{1}{3}t^{-2/3}\gen u,
\quad X_3=t\gen t+\frac{x}{3}\gen x-\frac{2}{3}u\gen u.$$
With the equivalence transformation
$$\tilde{t}=t,\quad \tilde{x}=x,\quad \tilde{u}=-3t^{2/3}u,$$
the basis elements are transformed, up to  scaling, to the
realization $A_{3,2}^3$. The transformed equation is
$$\tilde{u}_t=\tilde{u}_{xxx}-\frac{1}{3}t^{-2/3}\tilde{u}\tilde{u}_x.$$
This equation belongs to the class corresponding to the
realization $A_{3,2}^3$.

We note that a member of \eqref{vcKdV} for $f=1,$ $g=t^2$ (see
\cite{Gazeau92}) is equivalent, under appropriate point transformation, to
the above equation. Similarly, the particular case
$a=-\alpha/(1+\alpha)$, $\alpha\ne 0,1,2$ is equivalent to $f=1$,
$g=t^\alpha$ of \eqref{vcKdV}. In this case, the symmetry algebra
is indecomposable and solvable.

2.) $(a,k)=(-1,-1)$: the spherical KdV (sKdV) equation

In this case the equation is invariant with respect to a three-dimensional
symmetry algebra. We choose its basis to be
\begin{equation}
X_1=\gen x,\quad X_2=\ln t\gen x-\frac{1}{t}\gen u,\quad X_3=t\gen
t+\frac{x}{3}\gen x-\frac{2}{3}u\gen u,
\end{equation}
with non-zero commutation relations
$$[X_3,X_1]=-\frac{1}{3}X_1,\quad  [X_3,X_2]=X_1-\frac{1}{3}X_2.$$
It is easy to see that this algebra is isomorphic to  $A_{3,4}$.
Under the transformation $\tilde{u}=3tu$, the generators are
transformed to the realization $A_{3,4}^2$.  The sKdV eq. takes
the form
$$\tilde{u}_t=\tilde{u}_{xxx}+\frac{1}{3t}\tilde{u}\tilde{u}_x$$
which is a particular case of the equation invariant under the
algebra $A_{3,4}^2$.

We note that a member of \eqref{vcKdV} for $f=1,$ $g=e^{3t}$
\cite{Gazeau92} is equivalent to the case 2.), i.e. the sKdV
equation.

3.) $(a,k)=(a,0)$: The basis of the symmetry algebra reads as
\begin{equation}
X_1=\gen x,\quad \quad X_2=e^{at}(\gen x-a\gen u),\quad X_3=\gen
t,
\end{equation}
with non-zero commutation relation $[X_3,X_2]=a X_2$. The algebra
is isomorphic to $A_{3,2}$. With the transformation
$\tilde{u}=e^{-at}u$ the equation is transformed to a special case
of \eqref{vcKdV} for $f=1, g=e^{at}.$

4.) $(a,k)=(-1/2,-1)$: the cylindrical KdV (cKdV) equation

In this case the symmetry algebra is four-dimensional. In a convenient
basis we have
\begin{equation}\label{cKdValgebra}
\begin{gathered}
  X_1  = 2\sqrt t \partial _x  - \frac{1}
{{\sqrt t }}\partial _u , \hfill \\
  X_2  = 4t^{3/2} \partial _t  + 2x\sqrt t \partial _x  - (\frac{x}
{{\sqrt t }} + 4\sqrt t u)\partial _u , \hfill \\
  X_3  = \partial _x   ,\quad X_4  = 3t\partial _t  + x\partial _x  -
2u\partial _u
   \hfill \\
\end{gathered}
\end{equation}
with    non-zero commutation relations
$$[X_2,X_3]=-X_1,\quad [X_1,X_4]=-\frac{1}{2}X_1,\quad
[X_2,X_4]=-\frac{3}{2}X_2,\quad [X_3,X_4]=X_3.$$

We see that the symmetry  algebra of cKdV equation is isomorphic
to  the  algebra $A_{4,8}$ with $q=1$. The existence of such an
isomorphism is a necessary, but not sufficient condition for a
local point transforation to exist, transforming the two equations
into each other. Comparing these generators with \eqref{trvf} and
choosing \eqref{eq} suitably, for example, first transforming the
commuting elements $\curl{X_1, X_3}$ into $\{\partial_
{\tilde{x}}, \tilde{t}\partial_ {\tilde{x}}+\partial_
{\tilde{u}}\}$ and then transforming the remaining ones with the
aid of the freedom left in \eq{} we arrive at
$$\tilde{t}=2t^{-1/2},\quad \tilde{x}=t^{-1/2}x,\quad
\tilde{u}=tu+\frac{x}{2}$$ which establishes the equivalence of
the Lie algebra with basis \eqref{cKdValgebra} and cKdV equation
to  the KdV algebra ($A_{4,8}^1$) and KdV
equation. This connection between the KdV and cKdV equations is
well-known in the literature \cite{Bluman89}.

As a further comparison of the results obtained in the article we
consider
\begin{equation}\label{gkdv}
u_t+u_{xxx}+f(u)u_x^k=0,\quad k>0
\end{equation}
which is clearly a special case of \eqref{main}. Group
classification of this equation is given in a table (see Table
\ref{tab2}) \cite{Olver96}. These results can  immediately  be
derived from those obtained in this paper either directly or
performing a change of independent or dependent variables.

\begin{table}\caption{Symmetry Classification of \eqref{gkdv}}\label{tab2}
\begin{center}
\begin{tabular}{|c|c|c|l|c|}
\hline
$N$ & $k$  & $f(u)$ &     Symmetry  Generators             & Symmetry
Algebra \\
\hline
  1  & arb. &  arb.  &      $\gen t, \gen x$      &       $A_{2,1}^1$
\\
\hline
  2  & $k$  & $u^n$  & $\gen t, \gen x, t\gen t+\frac{x}{3}\gen
x+\frac{k-3}{k+n-1}u\gen u$,\quad $k+n\ne 1$ &     $A_{3,7}^2$     \\
\hline
  3  & $k$      & $e^u$       & $\gen t, \gen x, t\gen t+\frac{x}{3}\gen
x+(k-3)\gen u $  &  $A_{3,7}^2$       \\
\hline
  4  &  $k$    &  1      &  $\gen t, \gen x, t\gen t+\frac{x}{3}\gen
x+\frac{k-3}{3(k-1)}u\gen u, \gen u$,\quad  $k\ne 1$ &  $A_{4,3}^1$  \\
\hline
  5   &  3    &  arb.      &  $\gen t, \gen x, t\gen t+\frac{x}{3}\gen x$
  &    $A_{3,7}^1$              \\
\hline
  6   &  3    &   $u^{-2}$     &  $\gen t, \gen x, t\gen t+\frac{x}{3}\gen
x, u\gen u$     &  $A_{3,7}^1\oplus A_1$     \\
\hline
  7   &  3    &   1     &  $\gen t, \gen x, t\gen t+\frac{x}{3}\gen x, \gen
u$      &  $A_{3,7}^1\oplus A_1$ \\
\hline
  8   &  1    &  $u^n+c$      & $\gen t, \gen x, t\gen t+\frac{x}{3}\gen
x-\frac{2}{n}u\gen u$,\quad $n\ne 0$ & $A_{3,7}^2$  \\
\hline
  9   &  1    &    $u$    & $\gen t, \gen x, t\gen t+\frac{x}{3}\gen
x-\frac{2}{3}u\gen u, t\gen x+\gen u$   & $A_{4,6}^1$  \\
\hline
  10  &   1   &   $e^u+c$     &   $\gen t, \gen x, t\gen t+\frac{1}{3}(\gen
x+2ct)\gen x-\frac{2}{3}\gen u$  & $A_{3,7}^2$    \\
\hline
  11   &  1    &  1      & $\gen t, \gen x, t\gen t+\frac{1}{3}(x+2t)\gen x,
u\gen u, g(x,t)\gen u$      & Linear Eq.  \\
&&& $g_t+g_{xxx}+g_x=0$ & ($N=6$ in Table \ref{tab1})\\
\hline
\end{tabular}
\end{center}
\end{table}

Note that the equations that do not appear in the classification
list can be recovered from those by suitable point transformations.


A number of integrable KdV type equations can be reproduced by
restricting the arbitrary functions contained in invariant
equations of this article. For example, the realization
$A_{3,7}^2$ is equivalent to $\curl{\gen t, \gen x, t\gen t+x/3
\gen x-u/3\gen u}$ under the transformation $u\to u^{-1/3}$. We
have the invariant function
$$F=u^4 \tilde{F}(\omega_1,\omega_2), \quad \omega_1=u^{-2}u_x,\quad
\omega_2=u^{-3}u_{xx}.$$ Setting $\tilde F=\omega_1$ produces the
modified KdV (mKdV) equation
\begin{equation}\label{mkdv}
u_t=u_{xxx}+u^2 u_x.
\end{equation}
Since the maximal symmetry algebra of mKdV equation is
three-dimensional, it is not isomorphic to the KdV algebra. This
implies that there is no point transformation transforming
the mKdV equation into KdV equation. In this respect let us mention
that there is the well-known non-local transformation (Miura
transformation)
$$\tilde{u}=u^2\pm \sqrt{6}i u_x$$
taking the mKdV \eqref{mkdv} into the KdV equation
$\tilde{u}_t=\tilde{u}_{xxx}+\tilde{u}\tilde{u}_x$. Another
integrable equation which can be obtained from our classification
is \cite{Calogero}
$$u_t=u_{xxx}+3(u_{xx}u^2+3uu_{x}^2)+3u^4 u_x.$$
Its symmetry algebra is isomorphic to $A_{3,7}^2$.  We mention
that this equation can be linearized by a change of dependent
variable.

Let us mention also that a classification based on higher order
symmetries of third-order integrable nonlinear equations of the
form
\begin{equation}\label{high}
u_t=u_{xxx}+F(u,u_x,u_{xx})
\end{equation}
is given in \cite{Sokolov84}.

Finally, let us point out that in a very recent work
\cite{Peterson03} a class of integrable (in the sense of existence
of an infinite number of generalized symmetries) third-order
evolution equations of the form \eqref{high} for specific $F$
admitting recursion operators have been analyzed. Among others,
the special cases corresponding to $\tilde F=3\omega$ and $\tilde
F=3/2-3\omega$ of \eqref{inv} produce the following equations with
4-dimensional symmetry algebra $A_{3,7}^1\oplus A_1$
\[
\begin{gathered}
  u_t  = u_{xxx}  + 3u^{ - 1} u_x u_{xx} , \hfill \\
  u_t  = u_{xxx}  - 3u^{ - 1} u_x u_{xx}  + \frac{3}
{2}u^{ - 2} u_x ^3 , \hfill \\
\end{gathered}
\]
both of which were shown to admit recursion operators. This fact
indicates that many equations with relatively large symmetry
groups in our classification are among the most probable
candidates for being integrable.

We note that the maximal symmetry algebra of the first equation of
the above list is infinite-dimensional with basis elements
$$ X_1=\gen t,\quad X_2=\gen x,\quad
X_3=t\gen t+\frac{x}{3}\gen x+\frac{u}{2}\gen u,$$
$$X(\rho)=\rho(x,t)u^{-1}\gen u,\quad \rho_t=\rho_{xxx}.$$
The existence of an infinite-dimensional symmetry algebra suggests
linearizability of the equation by point transformations and,
indeed, it is linearized by the change of dependent variable
$v(x,t) = u^2(x,t)/ 2$.


\end{document}